\documentclass[usenatbib]{mn2e}

\usepackage{savesym}
\usepackage{amsmath}
\savesymbol{iint}
\usepackage{txfonts}
\usepackage{deluxetable}
\usepackage{html}

\usepackage{subfigure}
\restoresymbol{TXF}{iint}

\usepackage{color}
\usepackage[normalem]{ulem}

\usepackage{subfigure}
\usepackage{graphicx}
\usepackage{epstopdf}
\usepackage{url}

\def\simless{\mathbin{\lower 3pt\hbox
{$\rlap{\raise 5pt\hbox{$\char'074$}}\mathchar"7218$}}}   
\def\simmore{\mathbin{\lower 3pt\hbox
{$\rlap{\raise 5pt\hbox{$\char'076$}}\mathchar"7218$}}}   
\newcommand{\be}{\begin{equation}}
\newcommand{\ee}{\end{equation}}
\newcommand       \bea          {\begin{eqnarray}}
\newcommand       \eea          {\end{eqnarray}}
\newcommand       \apj          {ApJ}
\newcommand       \apjl         {ApJL}
\newcommand       \aap          {A\&A}
\newcommand       \nat          {Nature}
\newcommand       \mnras        {MNRAS}

\newcommand       \aj      {AJ}
\newcommand       \prd      {Phys.~Rev.~D.~}

\newcommand       \araa      {ARA\&A}

\newcommand       \jcap   {JCAP}

\newcommand      \apss {APSS}

\def\simlt{\mathrel{\hbox{\rlap{\hbox{\lower4pt\hbox{$\sim$}}}\hbox{$<$}}}}
\def\simgt{\mathrel{\hbox{\rlap{\hbox{\lower4pt\hbox{$\sim$}}}\hbox{$>$}}}}
\def\lesssim{\mathrel{\hbox{\rlap{\hbox{\lower4pt\hbox{$\sim$}}}\hbox{$<$}}}}

\def\gtrsim{\mathrel{\hbox{\rlap{\hbox{\lower4pt\hbox{$\sim$}}}\hbox{$>$}}}}

\title[Novae as Tevatrons]{Novae as Tevatrons: Prospects for CTA and IceCube}

\author[]{B.~D.~Metzger$^{1}\thanks{E-mail: bmetzger@phys.columbia.edu}$, D.~Caprioli$^{2}$, I.~Vurm$^{1}$, A.~M.~Beloborodov$^{1}$, I.~Bartos$^{1}$, A.~Vlasov$^{1}$\\
$^{1}$Columbia Astrophysics Laboratory, Columbia University, New York, NY 10027, USA\\
$^{2}$Department of Astrophysical Sciences, Princeton University, Princeton, NJ 08554, USA\\
}

\begin{document}
\date{Received / Accepted}
\pagerange{\pageref{firstpage}--\pageref{lastpage}} \pubyear{2014}

\maketitle

\label{firstpage}

\begin{abstract}
The discovery of novae as sources of $\sim 0.1-1$ GeV gamma-rays highlights the key role of shocks and relativistic particle acceleration in these transient systems.  Although there is evidence for a spectral cut-off above energies $\sim 1-100$ GeV at particular epochs in some novae, the maximum particle energy achieved in these accelerators has remained an open question.  The high densities of the nova ejecta ($\sim 10$ orders of magnitude larger than in supernova remnants) render the gas far upstream of the shock neutral and shielded from ionizing radiation.  The amplification of the magnetic field needed for diffusive shock acceleration requires ionized gas, thus confining the acceleration process to a narrow photo-ionized layer immediately ahead of the shock.  Based on the growth rate in this layer of the hybrid non-resonant cosmic ray current-driven instability (considering also ion-neutral damping), we quantify the maximum particle energy, $E_{\rm max}$, across the range of shock velocities and upstream densities of interest.  We find values of $E_{\rm max} \sim 10$ GeV - 10 TeV, which are broadly consistent with the inferred spectral cut-offs, but which could also in principle lead to emission extending to higher energies $\gtrsim 100$ GeV accessible to atmosphere Cherenkov telescopes, such as the planned Cherenkov Telescope Array (CTA).  Detecting TeV neutrinos with IceCube in hadronic scenarios appears to be more challenging, although the prospects are improved for a particularly nearby event (distance $\lesssim$ kpc) or if the shock power during the earliest, densest phases of the nova outburst is higher than is implied by the observed GeV light curves, due to downscattering of the gamma-rays by electrons within the ejecta.  Novae provide ideal nearby laboratories to study magnetic field amplification and the onset of cosmic ray acceleration, because other time-dependent sources (e.g. radio supernovae) typically occur too distant to detect as gamma-ray sources.  
\end{abstract} 
  

\section{Introduction}

Novae are sudden visual outbursts powered by runaway nuclear burning
on the surface of a white dwarf accreting from a stellar companion (e.g.~\citealt{Starrfield+98}).  They reach peak luminosities of $\sim 10^{4}L_{\odot}$ and
eject moderate quantities of mass $\sim 10^{-5}-10^{-4}M_{\odot}$ at
velocities of hundreds to thousands of km s$^{-1}$ (e.g., \citealt{Shore12}).  Although in standard models novae are thought to be powered directly by the
energy released from nuclear burning  (e.g.~\citealt{Hillman+14}), growing evidence suggests that
shock interaction plays an important role in powering nova emission
across the electromagnetic spectrum.  Evidence for shocks includes multiple velocity components in the optical spectra (e.g.~\citealt{Williams&Mason10}), hard X-ray emission starting weeks to years after the outburst (e.g.~\citealt{Mukai+08}; \citealt{Osborne+15}), and an early sharp maximum in the radio light curve on timescales of months, in excess of that expected from freely-expanding photo-ionized ejecta (e.g., ~\citealt{Chomiuk+14}; \citealt{Weston+15}).

\begin{figure*}
\includegraphics[width=1.0\textwidth]{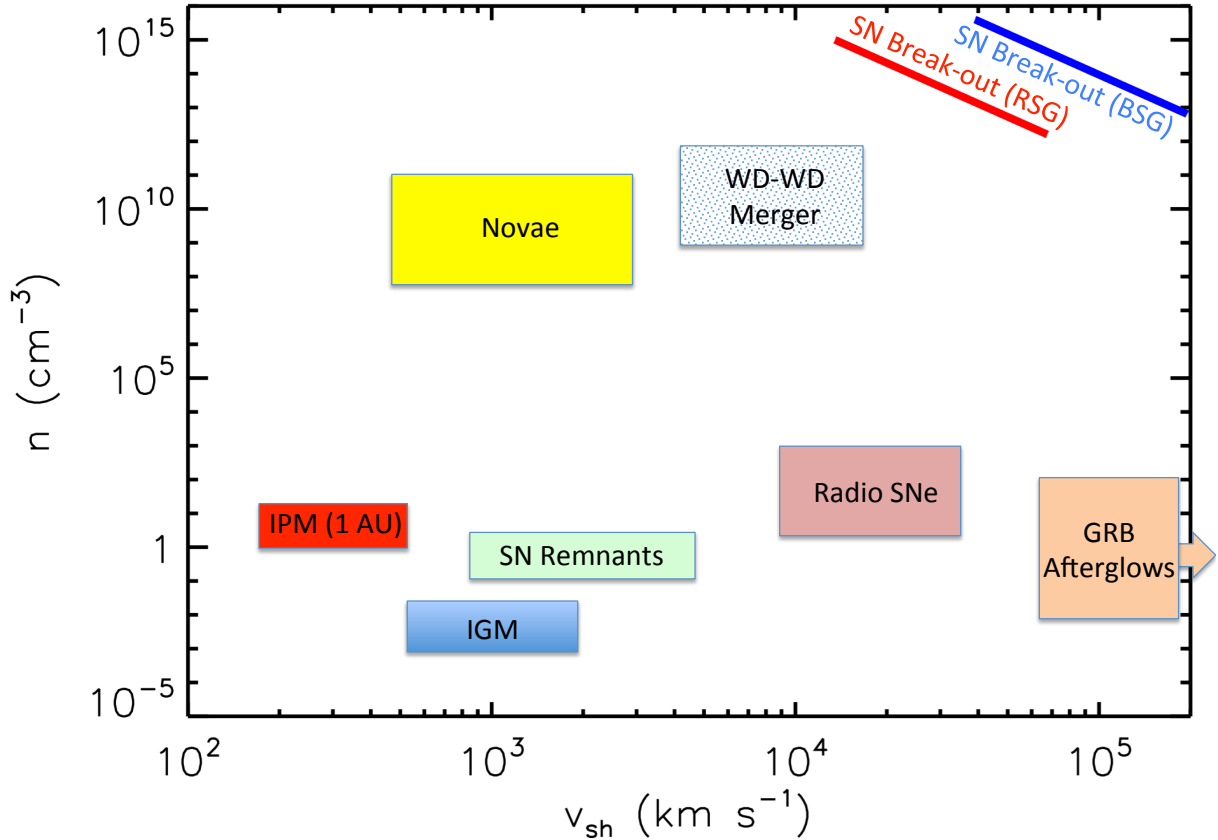}
\caption{Regimes of shock conditions in different astrophysical environments, in the space of shock velocity $v_{\rm sh}$ and density $n$.  Shocks in novae occupy a distinct portion of parameter space of relatively low shock velocities and high densities.  Shown for comparison are supernova shock break-out from red supergiant (RSG) and blue supergiant (BSG) progenitor stars, gamma-ray burst (GRB) afterglows, radio supernovae, supernova remnants, the intergalactic medium (IGM), interplanetary medium (IPM) on scales of 1 AU, and shocks in the outflows from binary white dwarf mergers (\citealt{Beloborodov14}).   } 
\label{fig:regimes}
\end{figure*}

The most striking indicator of shocks in novae is the recent discovery
by {\it Fermi} LAT of $\gtrsim$ 100 MeV gamma-rays, observed at times
coincident within a few days of the optical peak and lasting a few
weeks \citep{Ackermann+14}.  The first nova with detected gamma-rays occurred in the symbiotic binary V407 Cyg \citep{Abdo+10}, suggesting that shocks were produced by the interaction between the nova outflow and the dense wind of the companion red giant.  However, gamma-rays have now been detected from at least five ordinary classical novae with main sequence companions \citep{Ackermann+14}.  Remarkably, this demonstrates that the nova outflow runs into dense gas even in systems not embedded in the wind of an M giant or associated with recurrent novae.  This dense gas instead likely represents lower velocity mass ejected earlier in the outburst (`internal shocks'; \citealt{Mukai&Ishida01}, \citealt{Metzger+14}).  Current observations are consistent with many, and possibly all, novae producing shocks and $\gtrsim$ 100 MeV gamma-ray emission.  Indeed, the LAT-detected novae appear to be distinguished (if anything) only by their relatively close distances (e.g., \citealt{Finzell+15}).  Only relatively bright novae have triggered pointed {\it Fermi} observations, however, most of these resulting in detections.

Nova gamma-rays are produced by the decay of neutral pions created by proton-proton collisions, or via Inverse Compton scattering or bremsstrahlung emission from energetic electrons.  As both of these scenarios require a population of relativistic particles, these events provide a real-time probe of relativistic particle acceleration in non-relativistic shocks, complementary to those found in other astrophysical environments such as supernova remnants (e.g.~\citealt{Martin&Dubus13}; \citealt{Metzger+15}).  Constraints on the efficiency and energy spectra of particle acceleration as probed by novae provides a new and promising avenue to address old questions, such as the sources of Galactic cosmic rays and magnetic field amplification at shocks.  

Although the velocities of shocks in novae are similar to those in older supernova remnants, the density of the shocked gas in novae is typically 10 orders of magnitude higher (Fig.~\ref{fig:regimes}).  Such high densities result in several novel physical effects, such as the importance of radiative cooling on the post-shock dynamics (`radiative' shocks).  Furthermore, the gas both upstream well ahead of the shock and in the downstream cooling layer is largely neutral and hence will absorb thermal soft X-ray/UV radiation from the shock, reprocessing the shock power to optical frequencies (\citealt{Metzger+14}).  Indeed, a comparison between the minimum shock power needed to explain the observed gamma-ray luminosity and the optical light curve shows that a significant fraction of the latter is shocked-powered (\citealt{Metzger+15}), analagous to interacting supernovae (e.g.~\citealt{Chevalier&Irwin11}).  

The high densities of nova shocks also result in several physical simplifications.  The timescale for energy exchange between electrons and protons behind the shock via Coulomb scattering is short compared to the dynamical or thermal cooling rate, justifying a single temperature plasma (\citealt{Metzger+15}).  The high matter and photon densities near the shock furthermore ensure that relativistic particles radiate their energy rapidly compared to the outflow expansion time (fast cooling regime).  Gamma-ray producing novae therefore act as ``calorimeters" for probing the efficiency and spectrum of relativistic particle acceleration, in ways complementary to studies of supernova remnants (e.g.~\citealt{Morlino&Caprioli12}).  Exploiting the radiative nature of nova shocks, \citet{Metzger+15} place a minimum of $\epsilon_{\rm nth} \gtrsim 10^{-3}-10^{-2}$ on the fraction of the kinetic power of the shock placed into relativistic particles.  Based on observations of supernova remnant shocks (e.g.~\citealt{Volk+05}), and the results of MHD kinetic hybrid  (\citealt{Caprioli&Spitkovsky14}) and particle-in-cell (\citealt{Kato14}; \citealt{Park+14}) simulations, these high efficiencies appear to favor a hadronic origin for the GeV gamma-ray emission.  

This paper addresses a key question: what is the highest energy particle accelerated in nova shocks?  This issue is crucial to assessing novae as potential targets for present and future TeV telescopes, such as the Cherenkov Telescope Array (CTA; \citealt{Actis+11}), and current neutrino observatories, such as IceCube (\citealt{Karle+03}).  Although the statistics are poor, the time-integrated spectral energy distribution measured by {\it Fermi} LAT shows hints for a spectral cut-off or break in the photon energy range $E_{\gamma} \sim 1-10$ GeV (\citealt{Ackermann+14}).  \citet{MAGIC+15} presented upper limits on $\sim$ 0.1$-$1 TeV emission from V339 Del, which lie a factor of $\sim 10$ below the extrapolated {\it Fermi} LAT spectrum.  Similar TeV upper limits on the symbiotic nova V407 Cyg were reported by VERITAS \citep{Aliu+12}.  These observed cut-offs may result from an intrinsic steepening in the accelerated ion or electron spectrum, which in hadronic scenarios requires a break in the accelerated proton spectrum above $E \sim 10-100$ GeV.  Alternatively, an apparent cut-off may result from $\gamma-\gamma$ absorption due to $e^{\pm}$ pair-creation attenuation by the nova optical light.  This possibility is challenging to assess because the background photon energy density depends on the uncertain location where the shocks occur relative to the optical photosphere.

A naive application of the \citet{Hillas+84} confinement criterion, using the entire radial extent of the nova ejecta as the extent of the accelerator, results in a maximum particle energy of $\sim 1 $ PeV (\citealt{Metzger+15}).  However, this estimate is unrealistic because the magnetic field required for particle acceleration via diffusive shock acceleration cannot be supported in neutral regions (comprising the bulk of the ejecta) due to ion-neutral damping.  This limits the effective size of the acceleration zone to at most a narrow photo-ionized layer ahead of the shock, substantially reducing the maximum particle energy to a lower value (consistent with current observations).  Nevertheless, here we will show that particle energies exceeding $\sim 1$ TeV can readily be achieved under physically reasonable conditions during the nova outburst.    

This paper is organized as follows.  In $\S\ref{sec:shocks}$ we briefly review the properties of nova shocks to motivate the relevant physical conditions, including the radial extent of the photo-ionized layer.  In $\S\ref{sec:acceleration}$ we estimate the maximum particle energy in nova shocks by means of a simple stability analysis, based on the growth of the non-resonant \citet{Bell04} modes, considering also the role of ion-neutral damping.  In $\S\ref{sec:discussion}$ we discuss the implications of our results for novae as TeV photon (\S\ref{sec:Emax}) and neutrino (\S\ref{sec:neutrinos}) sources and for magnetic field amplification in non-relativistic shocks (\S\ref{sec:Bfield}).  We briefly summarize our results and conclude in $\S\ref{sec:conclusions}$.

\section{Nova Shocks}
\label{sec:shocks}

\begin{figure}
\includegraphics[width=0.5\textwidth]{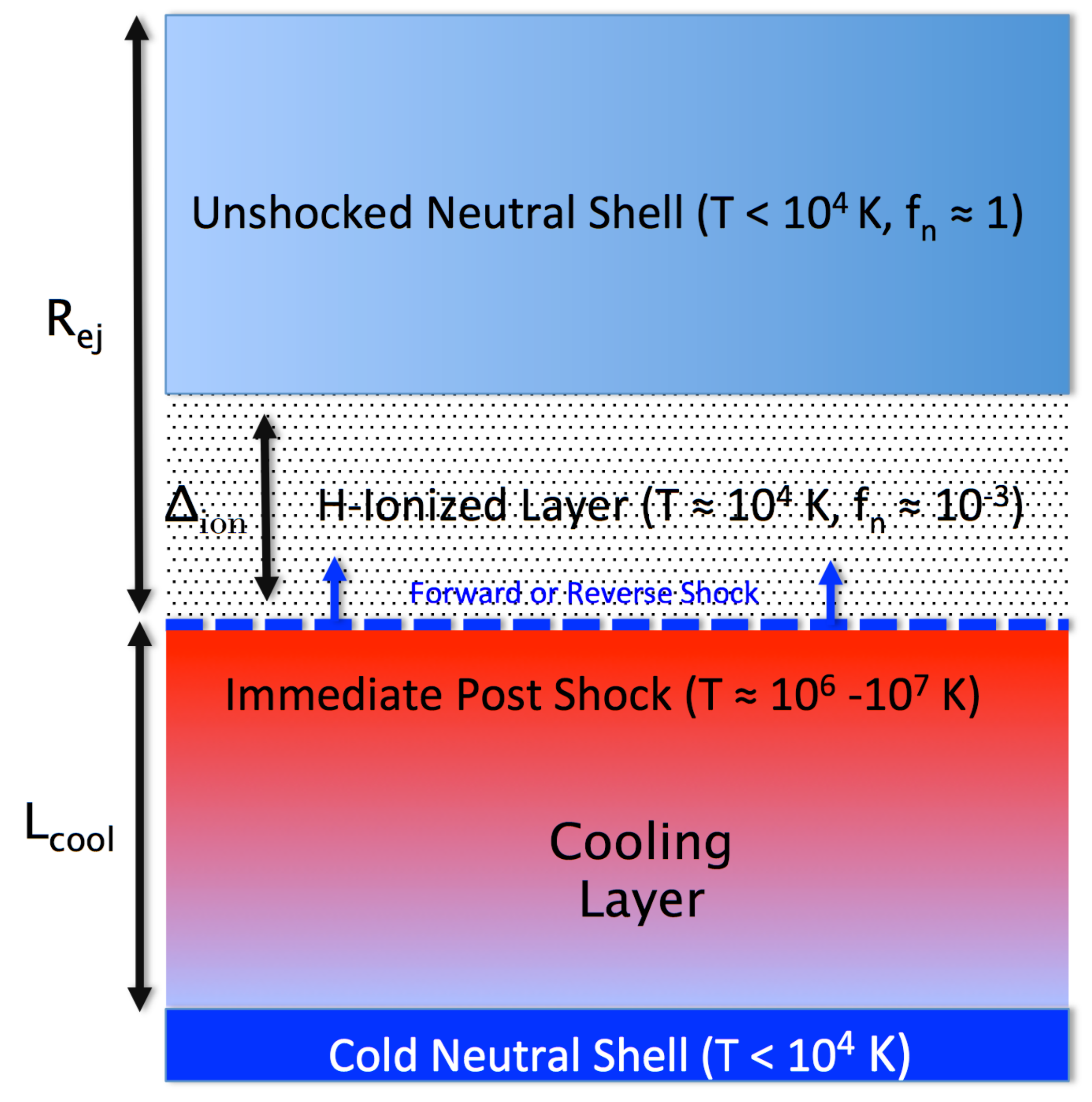}
\caption{Schematic diagram of shocks in novae, applicable to both forward and reverse shocks.  Gas well ahead of the shock is largely neutral ($f_n \approx 1$) due to the high recombination rate at high densities and low temperatures $T \lesssim 10^{4}$ K.  The post shock cools over a lengthscale $L_{\rm cool}$ (eq.~[\ref{eq:Lcool}]), which is typically smaller than the width of the preshocked shell (radiative shock).  X-ray/UV radiation from the shock photo-ionizes gas ahead of the shock, creating an ionized layer of thickness $\Delta_{\rm ion}$ (eqs.~[\ref{eq:deltaion}]) and temperature $T_{\rm us} \sim 1-2\times 10^{4}$ K in which $f_{\rm n} \approx 10^{-4}-10^{-3}$ (eq.~[\ref{eq:fn}]), the latter set by the balance between ionization and recombination.  Diffusive particle acceleration is limited to a fraction of the ionized layer $\lesssim \Delta_{\rm ion}$, which is itself generally much narrower than the ejecta thickness $R_{\rm ej}$ or post-shock cooling length $L_{\rm cool}$.} 
\label{fig:schematic}
\end{figure}

\subsection{Shock Dynamics}

Shocks were unexpected in classical novae because the pre-eruption environment surrounding the white dwarf is occupied only by the low density wind of the main sequence companion star, requiring a different source of matter into which the nova outflow collides.  One physical picture, consistent with both optical (e.g., \citealt{Schaefer+14}) and radio imaging (e.g., \citealt{Chomiuk+14}), and the evolution of optical spectral lines (e.g.,~\citealt{Ribeiro+13}; \citealt{Shore+13}), is that the thermonuclear runaway is first accompanied by a slow ejection of mass with a toroidal geometry, the shape of which may be influenced by the binary companion (e.g., \citealt{Livio+90}; \citealt{Lloyd+97}).  This slow outflow is then followed by a second discrete ejection or continuous wind (e.g., \citealt{Bath&Shaviv76}) with a higher velocity and more spherical geometry.  The subsequent collision between the fast and slow components produces strong ``internal" shocks within the ejecta which are concentrated in the equatorial plane.  The fast component continues to expand freely along the polar direction, creating a bipolar morphology (see Fig.~1 of \citealt{Metzger+15} for a schematic diagram).

The slow outflow of velocity $v_{\rm ej} = 10^{3}v_{\rm ej, 8}$ km s$^{-1}$ expands to a radius $R_{\rm ej} = v_{\rm ej} t \approx 6\times 10^{13}t_{\rm wk}v_{\rm ej,8}\,{\rm cm}$ by a time $t = t_{\rm wk}$ week.  The characteristic density of the ejecta of assumed thickness $\sim R_{\rm ej}$ can be estimated as
\be
n_{\rm ej} \approx \frac{M_{\rm ej}}{4\pi R_{\rm ej}^{3}f_{\Delta \Omega} m_p} \sim 9\times 10^{10}M_{-4}t_{\rm wk}^{-3}v_{\rm ej,8}^{-3}\,{\rm cm^{-3}},
\label{eq:nej}
\ee
where $f_{\Delta \Omega} \sim 0.5$ is the fraction of the total solid-angle subtended by the outflow and $M_{\rm ej} = 10^{-4}M_{-4}M_{\odot}$ is the ejecta mass, normalized to a characteristic value (e.g., \citealt{Seaquist+80}).  

A faster outflow (``wind") of mass loss rate $\dot{M}$ and velocity $v_{\rm f} \sim 2 v_{\rm ej}$ collides with the ejecta from behind.  The density of the wind at the collision radius ($\sim$ radius of the slow ejecta) is given by
\be
n_{\rm w} \approx \frac{\dot{M}}{4\pi R_{\rm ej}^{2}m_p v_{\rm f}} \sim 2\times 10^{9}\dot{M}_{-5}v_{\rm ej,8}^{-3}t_{\rm wk}^{-2}{\,\rm cm^{-3}},
\label{eq:nw}
\ee
where $\dot{M} = 10^{-5}\dot{M}_{-5}M_{\odot}\,{\rm wk^{-1}}$ is normalized to a value resulting in the ejection of $\sim 10^{-5}M_{\odot}$ over a week.  For instance, a total mass $4\times 10^{-5}M_{\odot}$ was ejected in the ``fast" component of V959 Mon (\citealt{Chomiuk+14}).  

The interaction drives a forward shock (FS) through the slow shell and a reverse shock (RS) back through the wind.  If the shocks are radiative, then the post shock material is compressed and piles up in a central cold shell sandwiched by the ram pressure of the two shocks (Fig.~\ref{fig:schematic}).  The FS propagates at a velocity $v_{\rm fs} = v_{\rm c}-v_{\rm ej} \ll v_{\rm ej}$, while the velocity of the reverse shock is $v_{\rm rs} = v_{\rm f}-v_{\rm c}$, where $v_{\rm c}$ is the velocity of the cold central shell, which only moderately exceeds the velocity of the slow ejecta due to the large inertia of the latter (see \citealt{Metzger+14}).  Hereafter, the velocities of both shocks are parametrized as $v_{\rm sh}= 10^{8}v_{8}$ cm s$^{-1} = \eta v_{\rm ej}$, where $\eta < 1$ for the FS and $\eta \sim 1$ for RS.  For parameters of interest, the RS dissipates kinetic energy at a comparable or greater rate than the FS, potentially favoring the former as the dominant site of particle acceleration.

The shocks heat the gas to a temperature
\be
T_{\rm sh} \simeq \frac{3}{16 k}\mu m_p v_{\rm sh}^{2} \approx 1.4\times 10^{7}v_{8}^{2}\,{\rm K},
\label{eq:Tsh}
\ee
and compresses it to a density $n_{\rm sh} = 4n$, where $\mu \approx 0.62$, $n = n_{\rm ej}$ (FS) or $n = n_{\rm w}$ (RS).  Gas behind the shock cools on a characteristic timescale
\begin{eqnarray}
&& t_{\rm cool}   = \frac{3kT_{\rm sh}/2\mu}{4n \Lambda(T_{\rm sh})} \approx
\left\{
\begin{array}{lr}
 2.0\times 10^{3} \eta v_{\rm ej, 8}^{4}M_{-4}^{-1}t_{\rm wk}^{3}\,\,{\rm s}
 &
{\rm FS}, \\
1.0\times 10^{5}  \eta v_{\rm ej,8}^{4}\dot{M}_{-5}^{-1}t_{\rm wk}^{2} \,\,{\rm s} &
{\rm RS}, \\
\end{array}
\right.
\label{eq:tcool}
\end{eqnarray}
where $\Lambda = \Lambda_0 T^{1/2} = 2\times 10^{-27}T^{1/2}$ erg cm$^{3}$ s$^{-1}$ is the free-free cooling function.  We neglect line cooling in this estimate, although it exceeds free-free cooling for $T_{\rm sh} \lesssim 3\times 10^{7}$ K ($v_{8} \lesssim 1$).  

Whether the shocks are radiative or adiabatic depend on whether the ratio of the cooling timescale to the expansion timescale,
\begin{eqnarray}
&&\chi \equiv \frac{t_{\rm cool}}{t} \approx
\left\{
\begin{array}{lr}
 3\times 10^{-3}\eta v_{\rm ej,8}^{4}M_{-4}^{-1}t_{\rm wk}^{2}
 &
{\rm FS}, \\
0.2  \eta v_{\rm ej,8}^{4}\dot{M}_{-5}^{-1}t_{\rm wk}&
{\rm RS}, \\
\end{array}
\right.
\label{eq:cool}
\end{eqnarray}
is less than unity.
Gas cools to low temperatures behind the shock on a lengthscale
\begin{eqnarray}
&&L_{\rm cool} \equiv v_{\rm sh}t_{\rm cool}/4 \approx
\left\{
\begin{array}{lr}
 5\times 10^{10} \eta^{2} v_{\rm ej,8}^{5}M_{-4}^{-1}t_{\rm wk}^{3}\,\,{\rm cm}
 &
{\rm FS}, \\
2\times 10^{12}  \eta v_{\rm ej,8}^{5}\dot{M}_{-5}^{-1}t_{\rm wk}^{2} \,\,{\rm cm} &
{\rm RS}, \\
\end{array}
\right.
\label{eq:Lcool}
\end{eqnarray}
such that the radiative shock condition can also be written as $L_{\rm cool} \ll R_{\rm ej}/4$.

The discussion above provides estimates for the characteristic range of gas densities $n \sim 10^{9}-10^{11}$ cm $^{-3}$ and shock velocities $v_{8} \sim 0.5-2$.  These values remain uncertain because the outflow and shock geometry are poorly understood theoretically and are observationally unresolved at the time of the gamma-ray emission.  However, the shock properties are constrained to produce the range of observed gamma-ray luminosities \citep{Ackermann+14}
\be
L_{\gamma} = \epsilon_{\rm nth}\epsilon_{\gamma}L_{\rm sh} \sim 10^{35-36} {\rm erg\,\, s^{-1}},
\label{eq:Lgamma}
\ee
where 
\be L_{\rm sh} = (9\pi/8)R_{\rm ej}^{2}n m_p v_{\rm sh}^{3} \sim 2\times 10^{38} n_{10}v_{8}^{5}t_{\rm wk}^{2}\,{\rm erg\,s^{-1}},
\label{eq:Lsh}
\ee 
is the kinetic power dissipated by the shocks, $n = 10^{10}n_{10}$ cm$^{-3}$ is the density of the upstream unshocked gas ($n_{\rm ej}$ or $n_w$ above), $\epsilon_{\rm nth} = 0.1\epsilon_{\rm nth,-1}$ is the fraction of the shock power used to accelerate relativistic non-thermal particles and $\epsilon_{\gamma} = 0.1\epsilon_{\gamma,-1}$ the fraction of this energy radiated in the LAT bandpass (\citealt{Metzger+15}).  In the second equality we have normalized the shock radius to a characterstic value of $R_{\rm ej} \sim v_{\rm sh}t$.

\subsection{Ionized Layer}

Absent external sources of photo-ionization, the unshocked upstream gas is neutral.  The recombination timescale $t_{\rm rec} \sim 1/n\alpha_{\rm rec}^{Z}$,
\be
\frac{t_{\rm rec}}{t} \sim 4\times 10^{-4}n_{10}^{-1}T_{\rm us,4}^{0.8}Z^{-2}t_{\rm wk}^{-1}
\ee
is short compared to the expansion time, where $\alpha_{\rm rec}^{Z} \approx 4\times 10^{-13}Z^{2}(T/10^{4}{\rm K})^{-0.8}$ cm$^{3}$ s$^{-1}$ is the approximate radiative recombination rate for hydrogen-like atomic species of charge $Z$ and $T_{\rm us} = T_{\rm us,4}10^{4}$ K is the temperature of the unshocked gas.  Absent photo-ionization heating, the shielded neutral gas will cool to a temperature $T_{\rm us} \lesssim$ few $10^{3}$ K, comparable or less than the effective temperature of the nova optical emission, for which $f_{\rm n} \sim \mathcal{O}(1)$ in thermal ionization balance (Saha equilibrium).

The upstream is subject to ionizing UV and X-ray radiation from the shocks, which penetrates the upstream medium to a depth $\Delta^{Z}$ (see eq.~[\ref{eq:deltaion}] below) that depends on the element $Z$ dominating the bound-free opacity at the frequency of relevance.  The neutral fraction of this exposed layer, $f_{\rm n,Z} = (1 + \lambda_{Z})^{-1}$, is set by the balance between ionization and recombination, where
\begin{eqnarray}
&&\lambda_{Z} \equiv \frac{4\pi}{\alpha_{\rm rec}^{Z}n_e}\int \frac{J_{\nu}}{h\nu}\sigma_{\rm bf}^{Z}(\nu)d\nu \approx \frac{L_{\rm sh}\sigma_{\rm thr}^{Z}}{12\pi \alpha_{\rm rec}^{Z} n R_{\rm sh}^{2}kT_{\rm sh}}\left(\frac{J_{\nu}}{J_{\rm f}}\right) \nonumber \\
&&\approx \frac{\sigma_{\rm thr}^{Z}v_{\rm sh}}{2\alpha_{\rm rec}^{Z}(1 + 5\chi/2)}\left(\frac{J_{\nu}}{J_{\rm f}}\right)
\sim  \frac{8\times 10^{2}v_{\rm 8}}{(1 + 5\chi/2)}Z^{-4}T_{\rm us,4}^{0.8}\left(\frac{J_{\nu}}{J_{\rm f}}\right).
\label{eq:fn}
\end{eqnarray}
Here $J_{\nu}$ is the average intensity of ionizing radiation, $n_e \sim n$ is the free electron density in the ionized layer (which is nearly fully ionized; see eq.~[\ref{eq:fn}] below), $\sigma_{\rm bf}^{Z} \simeq \sigma_{\rm thr}^{Z}(\nu/\nu_{\rm thr}^{Z})^{-3}$ is the approximate bound-free cross section, and $\sigma_{\rm thr}^{Z} \approx 6\times 10^{-18}Z^{-2}$ cm$^{2}$ is the approximate cross section at the ionization threshold frequency $h\nu_{\rm thr}^{Z} \approx 13.6Z^{2}$ eV (\citealt{Osterbrock&Ferland06}).  We have normalized $J_{\nu}$ to a value $J_{\rm f} \approx L_{\rm sh}h/16\pi^{2}R_{\rm sh}^{2}kT_{\rm sh}$ set by free-free cooling with its characteristic flat spectrum at frequencies $h\nu \ll kT_{\rm sh}$ (eq.~[\ref{eq:Tsh}]), neglecting line radiation and reprocessing (we revisit this assumption below).  The factor $(1 + 5\chi/2)^{-1}$ accounts for the radiative efficiency of the shock (eq.~[\ref{eq:cool}]; \citealt{Metzger+14}).  

Equation (\ref{eq:fn}) shows that $\lambda_{Z} \gg 1$ ($f_{\rm n,Z} \ll 1$) for light elements with $Z \lesssim 5$.  Radiation of frequency $\nu \sim \nu_{\rm thr}^{Z}$ penetrates the neutral gas to the depth $\Delta^{Z}$ where the absorptive optical depth equals unity, 
\be (n/A)(\sigma_{\rm thr}^{Z}/2)X_Z f_{\rm n,Z}\Delta^{Z} \sim  1,
\ee
where $X_Z$ is the mass fraction of element $Z$ and mass number $A$ and we have taken a characteristic cross section $\sigma_{\rm thr}^{Z}/2$ as half that of the threshold value (see Appendix A of \citealt{Metzger+14} for justification).  For $\Delta^{Z} \ll R_{\rm ej}$ we thus have (in the $\lambda_Z \gg 1$ limit)
\begin{eqnarray}
\Delta^{Z} \underset{\lambda_Z \gg 1}\approx  \frac{A(J_{\nu}/J_{\rm f})}{X_{Z}(1 + 5\chi/2)}\frac{v_{\rm sh}}{n\alpha_{\rm rec}^{Z}} 
\sim  \frac{3\times 10^{10}\,{\rm cm}}{1+5\chi/2}\frac{v_{8}}{X_Z}\frac{A}{Z^{2}}n_{\rm 10}^{-1}T_{\rm us,4}^{0.8}\left(\frac{J_{\nu}}{J_{\rm f}}\right)
\label{eq:deltaion}
\end{eqnarray}
where we have used equation (\ref{eq:fn}). Elements with higher $Z$ are ionized to a smaller depth $\Delta^{Z} \propto A/(X_{Z}Z^{2})$, so for characteristic nova abundances H-ionizing photons penetrate the deepest and determine the extent of the ionized layer:
\be
\Delta_{\rm ion} = \Delta^{Z = 1} \simeq 4\times 10^{10}\,{\rm cm}\frac{v_{8}T_{\rm us,4}^{0.8}}{(1+5\chi/2)n_{10}}\left(\frac{J_{\nu}}{J_{\rm f}}\right),
\label{eq:DeltaionH}
\ee
where we have taken $X_{Z = 1} = 0.5$.  This is typically 2$-$3 orders of magnitude smaller than the characteristic radius of the ejecta at the time of the shocks, $R_{\rm ej} \sim 10^{13}$ cm.  The residual neutral fraction in the ionizing layer is approximately
\be
f_{\rm n} \approx \lambda_{Z=1}^{-1} \approx 1.3\times 10^{-3}(1+5\chi/2)v_{8}^{-1}T_{\rm us,4}^{-0.8}\left(\frac{J_{\nu}}{J_{\rm f}}\right)^{-1}
\label{eq:fn}
\ee

Direct free-free shock luminosity sets the {\it minimum} ionizing intensity, but $J_{\nu}$ can exceed $J_{\rm f}$ if line cooling becomes more important than free-free emission or due to ionization by secondary photons produced by the absorption and reprocessing of higher frequency radiation by the neutral gas.  Such reprocessing is challenging to determine accurately without a detailed photo-ionization calculation, but a simple estimate suggests that the extent of the ionizing layer could be expanded by an order of magnitude or more.   If a fraction $f_{\rm EUV}$ of the total shock luminosity is placed into H-ionizing photons of energy $\sim 2h\nu_{\rm thr}^{Z=1}$, then the maximum thickness of the ionized layer is larger than the minimum value of $\Delta_{\rm ion}$ (eq.~[\ref{eq:deltaion}]) by a factor of 
\be
J_{\nu}/J_{\rm f} \sim 1 + f_{\rm EUV}(kT_{\rm sh}/2h\nu_{\rm thr}^{Z=1}) \approx 1 + 8(f_{\rm EUV}/0.1)v_{8}^{2}.  
\label{eq:Jnu}
\ee

\section{Particle Acceleration}
\label{sec:acceleration}

In the standard theory of Diffusive Shock Acceleration (DSA; e.g., \citealt{Blandford&Ostriker78}), particles are accelerated to energies exceeding that of the thermal plasma by diffusing back and forth across the shock front via interaction with magnetic turbulence upstream and downstream.  The momentum distribution, $f(p) \propto p^{-4}$, predicted by the simplest DSA models corresponds to an energy distribution
\begin{eqnarray}
&& \frac{dN}{dE}E^{2} \propto  
\left\{
\begin{array}{lr}
E^{1/2}
, &
kT_{\rm sh} \lesssim E \ll m_p c^{2} \\
{\rm constant,} &
m_p c^{2} \ll E < E_{\rm max}, \\
\end{array}
\right..
\label{eq:dNdEion}
\end{eqnarray}
that concentrates the non-thermal energy in relativistic particles.  Equation (\ref{eq:dNdEion}) is valid for strong shocks, which are justified because the post-shock temperature $T_{\rm sh}$ greatly exceeds the temperature of the upstream ionized gas, $T_{\rm us} \sim 10^{4}$ K.  Because of the high degree of ionization of the upstream gas (eq.~[\ref{eq:fn}]), the effects of a neutral return flux (e.g., \citealt{Blasi+12}) do not appear to be relevant to nova shocks.  

The magnetic field of the unshocked nova ejecta will likely be weak due to flux freezing dilution from the white dwarf surface (\citealt{Metzger+15}).  The confinement needed for relativistic particle acceleration in nova shocks thus requires significant magnetic field {\it amplification}.  A promising candidate for such amplication on small-scales in non-relativistic shocks is the hybrid non-resonant (NRH) cosmic ray current-driven instability identified by \citet{Bell04} (see also \citealt{Bell05}; \citealt{Reville+06}; \citealt{Amato&Blasi08}; \citealt{Zirakashvili+08}; \citealt{Riquelme&Spitkovsky09}; \citealt{Caprioli&Spitkovsky14b}).  However, the high densities in novae imply that ion-neutral collisions can potentially damp the NRH (\citealt{Reville+07}), which could suppress particle acceleration in the neutral or quasi-neutral regions ahead of the shock (Fig.~\ref{fig:schematic}).  

The growth rate of the NRH modes, accounting for ion-neutral damping, is given by (\citealt{Reville+07})
\be
\gamma = 
\left\{
\begin{array}{lr}
 \sqrt{\nu_{\rm in}^{2}/4 + \gamma_{\rm fi}^{2}} -\nu_{\rm in}/2 \underset{\nu_{\rm in} \gg \gamma_{\rm fi}}\approx \gamma_{\rm fi}^{2}/\nu_{\rm in}
 &
f_{\rm n} \sim 1, \\
\gamma_{\rm fi} &
f_{\rm n} \ll 1, \\
\end{array}
\right.
\label{eq:growth}
\ee  
where $\gamma_{\rm fi} \equiv (k \zeta v_{\rm sh}^{2}/r_{\rm g,min})^{1/2}$ is the growth rate for fully ionized gas and $\nu_{\rm in} \approx 90f_{\rm n}n_{10}T_{\rm us,4}^{0.4}{\rm s^{-1}}$
is the ion-neutral momentum exchange frequency (\citealt{Kulsrud&Cesarsky71}).  Here $\zeta \approx \epsilon_{\rm p}v_{\rm sh}/c$ is a dimensionless parameter characterizing the strength of the ion current driving term, which is proportional to the fraction $\epsilon_{\rm p}$ of the shock power used to accelerate relativistic protons, and $r_{\rm g,min} = p_{\rm min}c/eB$ is the  gyroradius of the minimum energy cosmic rays, where $p_{\rm min}$ is the minimum momentum of the current-carrying ions.  The simplification made in the top line of equation (\ref{eq:growth}) that $\nu_{\rm in} \gg \gamma_{\rm fi}$ has been checked after the fact.

The minimum cosmic ray momentum $p_{\rm min} \approx E_{\rm min}/c$ increases with distance ahead of the shock because particles with larger energy (larger gyroradii) can diffuse further in the face of downstream advection.  The minimum ion energy $E_{\rm min}(z)$ a distance $z$ ahead of the shock is estimated by equating the upstream diffusion timescale $t_{\rm diff} \sim D/v_{\rm sh}^{2}$ to the advection time $t_{\rm adv} = z/v_{\rm sh}$ from $z$ to the shock, where $D$ is the diffusion coefficient.  Taking $D = r_{\rm g}c/3$, corresponding to the limit of Bohm diffusion (\citealt{Caprioli&Spitkovsky14c}), where $r_{\rm g} = E/eB$ is the gyroradius and $B$ is the magnetic field strength, gives
\begin{eqnarray}
E_{\rm min}(z) &\approx& 3 eB v_{\rm sh}z/c \approx 170\,{\rm GeV}v_{8}^{2}\epsilon_{B,-2}^{1/2}n_{10}^{1/2}\left(\frac{z}{10^{10}\rm cm}\right) \nonumber \\ &\approx& \frac{700 {\rm\, GeV}}{1+5\chi/2}v_{8}^{3}\epsilon_{B,-2}^{1/2}n_{10}^{-1/2}T_{\rm us,4}^{0.8}\left(\frac{z}{\Delta_{\rm ion}}\right)\left(\frac{J_{\nu}}{J_{\rm f}}\right),
\label{eq:Emin}
\end{eqnarray}
where in the final line of equation (\ref{eq:Emin}), $z$ is normalized to the thickness of the H-ionized layer (eq.~[\ref{eq:DeltaionH}]).  We have normalized the magnetic field strength, $B$, to its post-shock value
\begin{eqnarray}
B_{\rm sh} &=& \sqrt{6\pi \epsilon_{B}m_p n v_{\rm sh}^{2}} = 5.6 v_{8}n_{10}^{1/2}\epsilon_{B,-2}^{1/2} {\rm G},
\label{eq:Bsh}
\end{eqnarray}
which is parameterized as a fraction $\epsilon_{B,-2} = \epsilon_{B}/0.01$ of the post-shock pressure $P = (3/4)m_p n v_{\rm sh}^{2}$.  Strictly speaking, the growth rate should be calculated using the initial, unamplified magnetic field strength $B_0$, which corresponds to $\epsilon_{B} \ll 0.01$.  However, a larger value of $\epsilon_{B} \sim 10^{-2}$ is motivated by simulations of the NRH, which show that in the non-linear stage the growth of the instability is {\it reduced} by a compensating factor of $(B/B_0)^{2}$ \citep{Caprioli&Spitkovsky14b}.  Thus, although our estimate may be wrong in the first e-fold of the growth, it becomes better and better and converges to the growth rate in the amplified field once saturation is achieved.

Using equation (\ref{eq:Emin}), the growth rate (eq.~[\ref{eq:growth}]) is now written
\be
\gamma \approx 
\left\{
\begin{array}{lr}
 kr_{\rm g}\cdot\epsilon_{\rm p}v_{\rm sh}^{2}eB/(3\nu_{\rm in}E z)
 &
f_{\rm n} \sim 1, \\
\left[kr_{\rm g}\cdot\epsilon_{\rm p}v_{\rm sh}^{2}eB/(3E z)\right]^{1/2} &
f_{\rm n} \ll 1.\\
\end{array}
\right.
\label{eq:growth2}
\ee  
Because the growth rate increases with $k$ and particles of energy $E$ scatter most effectively off modes of wavenumber $k \lesssim 1/r_{\rm g}$, the modes with $kr_{\rm g} \sim 1$ are the most relevant ones. Such modes only have time to act if their growth time, $t_{\rm gr} = 1/\gamma$, is less than the time it takes the disturbance to be advected downstream, $t_{\rm adv} = z/v_{\rm sh}$.  This ratio is given by
\begin{eqnarray}
&&\left.\frac{t_{\rm gr}}{t_{\rm adv}}\right|_{k r_{\rm g} =1} \approx \nonumber \\
&&\left\{
\begin{array}{lr}
 3\nu_{\rm in}E/(\epsilon_{\rm p}v_{\rm sh}eB) \approx 16f_{\rm n}\epsilon_{\rm p,-1}^{-1}\epsilon_{B,-2}^{-1/2}v_{8}^{-2}n_{10}^{1/2}T_{\rm us,4}^{0.4}\left(\frac{E}{{\rm GeV}}\right) &
f_{\rm n} \sim 1, \\
\sqrt{3E/(\epsilon_{\rm p}zeB)} &
f_{\rm n} \ll 1. \\
\end{array}
\right. \nonumber \\
\label{eq:growthratio}
\end{eqnarray}
Equation (\ref{eq:growthratio}) shows that in the far upstream neutral layer ($f_{\rm n} = 1$), we have $t_{\rm gr} \gg t_{\rm adv}$, making the acceleration of particles with energies $E \gg $ GeV unlikely in this region.  

However, the lower value of $f_{\rm n} \sim 10^{-3} \ll 1$ in the H-ionized layer (eq.~[\ref{eq:fn}]) allows for more efficient particle acceleration.  Because $t_{\rm gr}/t_{\rm adv}$ decreases with distance from the shock, the maximum energy particle is limited by the growth rate at $z \sim \Delta_{\rm ion}$.  At this location, we have $t_{\rm gr} \lesssim t_{\rm adv}$ for particles up to an energy
\begin{eqnarray}
E_{\rm max, 1} &\approx& \left.\frac{\epsilon_{\rm p}z eB}{3}\right|_{z = \Delta_{\rm ion}} \approx 2.2{\,\rm TeV}\frac{\epsilon_{\rm p,-1}v_{8}^{2}\epsilon_{B,-2}^{1/2}T_{\rm us,4}^{0.8}}{n_{10}^{1/2}(1 + 5\chi/2)}\left(\frac{J_{\nu}}{J_{\rm f}}\right).
\label{eq:Emax1}
\end{eqnarray}
Equation (\ref{eq:Emax1}) sets the maximum particle energy only when $E_{\rm max,1}$ is less than the value limited by the size of the ionized layer (diffusive confinement criterion), which from equation (\ref{eq:Emin}) is given by
\begin{eqnarray}
E_{\rm max,2} = E_{\rm min}(z = \Delta_{\rm ion}^{\rm min}) \approx \frac{700{\rm \,GeV}}{(1+5\chi/2)}v_{8}^{3}\epsilon_{B,-2}^{1/2}n_{10}^{-1/2}T_{\rm us,4}^{0.8}\left(\frac{J_{\nu}}{J_{\rm f}}\right).
\label{eq:Emax2}
\end{eqnarray}
The fact that $E_{\rm max,2} = [9v_{\rm sh}/(\epsilon_{p}c)]E_{\rm max,1} \approx 0.3v_{8}\epsilon_{p,-1}^{-1}E_{\rm max,1}$ shows that the thickness of the ionized region is generally more important than the finite NRH growth rate for fiducial parameters.  $E_{\rm max,1}$ and $E_{\rm max,2}$ are increasing functions of both the shock velocity $v_{8}$ and the amount of ionizing radiation $J_{\nu}/J_{\rm f}$, the latter of which is expected to depend inversely on $v_{8}$ due to the greater importance of line emission and radiation trapping/downscattering for lower shock velocities.


Finally, we check that the cosmic ray acceleration time $t_{\rm acc} \approx 8D/v_{\rm sh}^{2}$ (e.g., \citealt{Caprioli&Spitkovsky14c}), 
\begin{eqnarray}
t_{\rm acc} \approx \frac{Ec}{3eB v_{\rm sh}^{2}} \underset{\rm eq.~[\ref{eq:Bsh}]}\approx  10^{-3}{\rm week}\,\, v_{\rm 8}^{-3}\epsilon_{B,-2}^{-1/2}n_{10}^{-1/2}(E/{\rm TeV}),
\end{eqnarray}
 is much less than expansion time $t$.  The acceleration time is also shorter than the timescale for pionic losses, $t_{\pi} \sim 1/(4n\sigma_{\rm \pi}c)$,
\be \frac{t_{\rm acc}}{t_{\pi}} \approx   0.014 v_{\rm 8}^{-3}\epsilon_{B,-2}^{-1/2}n_{10}^{1/2}t_{\rm wk}^{-1}(E/{\rm TeV}),
\ee
where we have taken $\sigma_{\pi} \approx 20$ mbarn as an estimate of the p-p inelastic cross section (\citealt{Kamae+06}).  A similar calculation shows that Coulomb losses are unimportant on the acceleration timescale.  The maximum particle energy is thus set by the confinement condition, rather than the finite acceleration time.


\section{Implications}
\label{sec:discussion}

\subsection{Maximum Particle Energy}
\label{sec:Emax}

\begin{figure}
\includegraphics[width=0.5\textwidth]{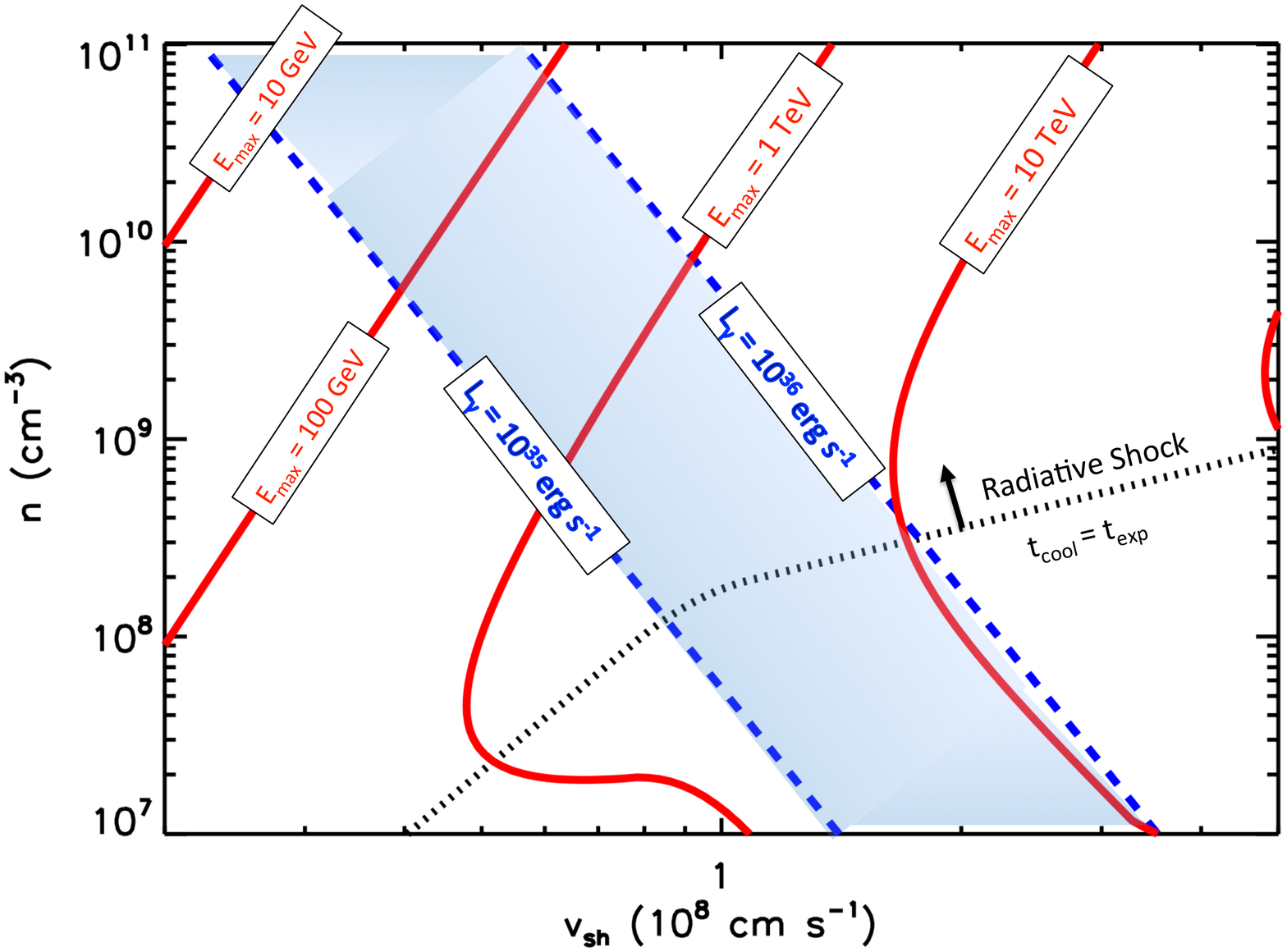}
\includegraphics[width=0.5\textwidth]{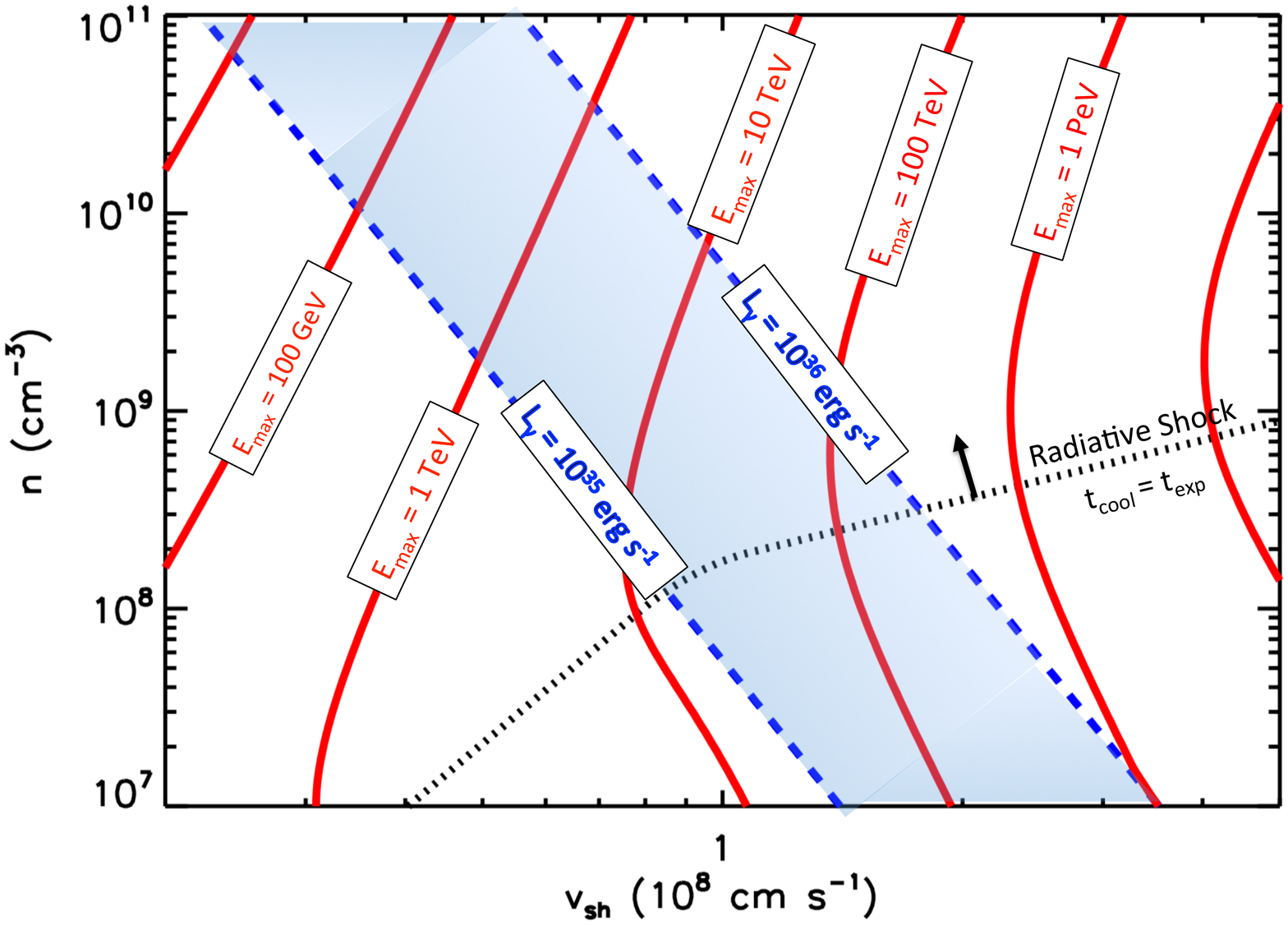}
\caption{Maximum particle energy (solid red lines) accelerated at the shock as a function of the shock velocity $v_{\rm sh}$ and upstream density $n$, calculated as the minimum of $E_{\rm max,1}$ (eq.~[\ref{eq:Emax1}]) and $E_{\rm max,2}$ (eq.~[\ref{eq:Emax2}]) for $\epsilon_{B} = 10^{-2}$ and shown for two cases: assuming (1) {\bf Top Panel:} photo-ionization exclusively from the shock's free-free emission $J_{\nu} = J_{\rm f}$, (2) {\bf Bottom Panel}: assuming a fraction $f_{\rm EUV} = 0.1$ of the total shock power is available to ionize hydrogen via direct line emission or reprocessing of hard radiation absorbed by neutral gas (eq.~[\ref{eq:Jnu}]).  A dashed black line separates the regime of adiabatic versus radiative shocks ($t_{\rm cool} = t_{\rm exp}$), where $t_{\rm exp} = 2$ weeks is the characteristic timescale of the LAT emission and $t_{\rm cool}$ is calculated using a standard cooling function for solar metallicity gas (\citealt{Schure+09}).  Dashed blue lines show the gamma-ray luminosity $L_{\gamma}$ (eq.~[\ref{eq:Lgamma}]), calculated assuming a shock radius $R_{\rm ej} = v_{\rm sh}t$, $\epsilon_{\rm nth} = 0.1$, and $\epsilon_{\gamma} = 0.1$ (\citealt{Metzger+15}), bracketing the range $L_{\gamma} \sim 10^{35}-10^{36}$ erg s$^{-1}$ measured by {\it Fermi} LAT.  The largest values of $E_{\rm max}$ at high $v_{8}$ in the bottom panel may be unphysical because the $\gtrsim$ keV thermal X-rays from such high velocity shocks will directly escape instead of ionizing the upstream gas.  } 
\label{fig:Emax1}
\end{figure}

Figure \ref{fig:Emax1} shows contours of the maximum particle energy $E_{\rm max} =$ min$[E_{\rm max,1},E_{\rm max,2}]$ in the space of the shock velocity $v_{\rm sh}$ and upstream density $n$.  The top panel is calculated assuming photo-ionization ahead of the shock exclusively from the shock's free-free emission $J_{\nu} = J_{\rm f}$, while the bottom panel is calculated assuming that an additional fraction $f_{\rm EUV} = 0.1$ of the total shock power is available to ionize hydrogen, either via direct line emission or reprocessing of hard radiation absorbed by neutral gas (eq.~[\ref{eq:Jnu}]).  The blue region shows the parameter space of shock power necessary to explain the observed gamma-ray luminosities, $L_{\gamma} \sim 10^{35}-10^{36}$ erg s$^{-1}$ (eq.~[\ref{eq:Lgamma}]), calculated assuming a shock radius $R_{\rm ej} = v_{\rm sh}t$, $\epsilon_{\rm nth} = 0.1$, and $\epsilon_{\gamma} = 0.1$ (\citealt{Metzger+15}).  

\begin{figure}
\includegraphics[width=0.5\textwidth]{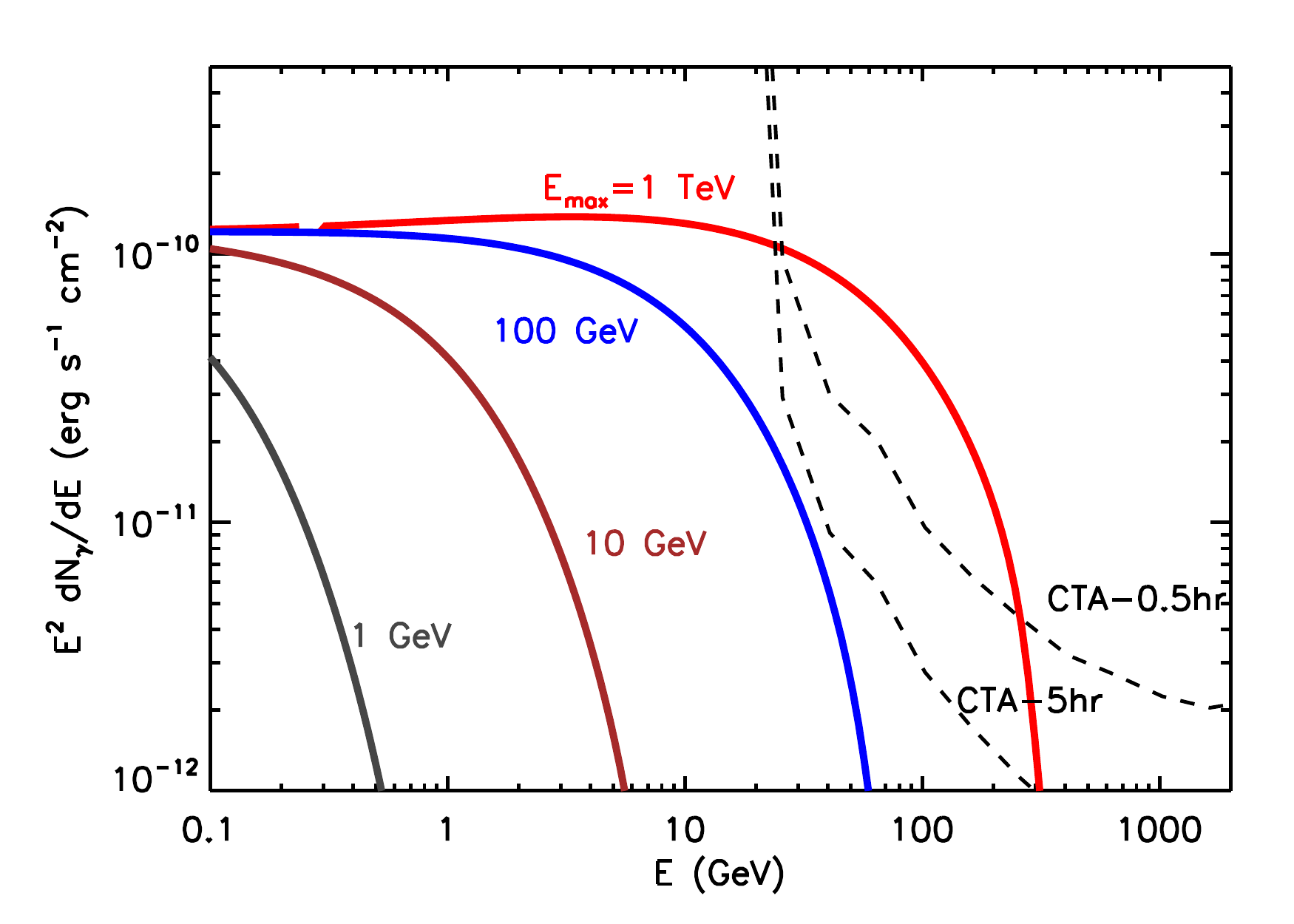}
\caption{Spectral energy distributions from pion decay \citep{Kamae+06}, calculated assuming the DSA particle spectrum (eq.~[\ref{eq:dNdEion}]) for different assumed values for the spectral cut-off as marked.  The 0.1-1 GeV flux is normalized to characteristic values of novae detected by Fermi LAT \citep{Ackermann+14}.  Also shown for comparison are the predicted 5$\sigma$ sensitivity of the Cherenkov Telescope Array for 0.5 and 5 hour integrations, respectively.} 
\label{fig:spectra}
\end{figure}

Maximum particle energies ranging from $E_{\rm max} \sim 10$ GeV$-$10 TeV are achieved across the preferred parameter space ($n \sim 10^{8}-10^{11}$ cm$^{-3}$), with $E_{\rm max}$ larger for lower gas densities and/or higher shock velocities.  This range is broadly consistent with the range $\sim 10-100$ GeV needed to explain the observed spectral cut-offs in the LAT-detected novae, which for hadronic scenarios typically occur at roughly $E_{\rm \gamma,max} \sim E_{\rm max}/10$ (Fig.~ \ref{fig:spectra}).  Note also that $E_{\rm max} \propto \epsilon_{B}^{1/2}$ depends on the magnetic energy fraction, which is taken to be $\epsilon_{B} = 10^{-2}$ in Figure \ref{fig:Emax1} (\citealt{Caprioli&Spitkovsky14b}).

A spectral cut-off can also result if photons are attenuated due to $\gamma-\gamma$ absorption within the source by the nova optical light.  Photons of energy 100 GeV can pair create off target photons of energy $E_{\rm opt} \sim 1$ eV, resulting in an optical depth
\begin{eqnarray}
\tau_{\gamma-\gamma} \approx n_{\gamma}\sigma_{\gamma-\gamma}R_{\rm ej} \approx 3 (\tau_{\rm opt}+1)\left(\frac{E_{\rm opt}}{\rm eV}\right)^{-1}\left(\frac{L_{\rm opt}}{10^{39}\,{\rm erg\,s^{-1}}}\right)v_{8}^{-1}t_{\rm wk}^{-1}
\label{eq:TeV}
\end{eqnarray}
near unity, where $\sigma_{\gamma-\gamma} \approx 10^{-25}$ cm$^{2}$ is the photon-photon absorption cross section near threshold and $n_{\gamma} \approx L_{\rm opt}(\tau_{\rm opt}+1)/(4\pi c R_{\rm ej}^{2}E_{\rm opt})$ is the energy density of the target optical photons near the shock and $\tau_{\rm opt} \simeq m_p n R_{\rm ej}\kappa_{\rm opt} \approx 0.1 v_{\rm ej,8} n_{10}t_{\rm wk}$ is the optical depth of the shock at optical frequencies, where $\kappa_{\rm opt} \sim 0.1$ cm$^{2}$ g$^{-1}$ is the optical opacity.  We find that $\tau_{\gamma-\gamma} \lesssim$ few on a timescale of weeks for the relevant range of $n-v_{8}$ parameter space where $E_{\rm max}$ is highest, suggesting that $\gamma-\gamma$ may be relevant.  However, due to the decreasing target photon number density from novae in the infrared (the Rayleigh Jeans tail, i.e. neglecting dust formation), such suppression may not extend to $\gtrsim 1$ TeV energies.

Although the {\it Fermi} LAT spectra hint at a $\sim 10$ GeV spectral cut-off \citep{Ackermann+14}, which was confirmed with high significance at a single observing epoch by MAGIC (\citealt{MAGIC+15}), our results in Figure \ref{fig:Emax1} suggest that this does not exclude novae as TeV sources.  The shock properties ($n$, $v_{\rm sh}$, $\tau_{\gamma-\gamma}$) are likely to vary in time during the nova outburst, so a spectral cut-off measured at one epoch does not exclude a higher value of $E_{\rm max}$ at other times. 

CTA will reach a 5$\sigma$ sensivity of $\sim 3\times 10^{-11}-10^{-11}$ erg cm$^{-1}$ s$^{-1}$ in the energy range of $\sim 0.1-1$ TeV for a half-hour integration, and a factor 3 times deeper for a 5 hour integration (\citealt{Actis+11}).  Figure \ref{fig:spectra} shows the spectral energy distribution from pion decay, assuming the DSA particle spectrum (eq.~[\ref{eq:dNdEion}]) for different values of the proton spectral cut-off and normalized to the characterstic value of the $0.1-1$ GeV nova LAT flux.  We find that a LAT-detected nova with $E_{\rm max} \gtrsim 1$ TeV could easily be detected by CTA in a half-hour integration, while a nova with the same $E_{\rm max}$ even an order of magnitude dimmer could be detected in 5 hours.  We strongly encourage future additional TeV follow-up of novae near and after optical peak, even if not first detected by {\it Fermi}.  Detections or upper limits can be used to place meaningful constraints on the maximum accelerated particle energy and the location of the gamma-ray emission.   

Although roughly 10 Galactic novae are discovered by optical surveys per year, only $\sim 1$ per year has been detected by {\it Fermi-}LAT.  The main distinguishing feature of LAT-detected novae appears to be their close distances, although selection effects related to which novae receive pointed observations cannot be excluded. The deeper sensivity of CTA suggests that, if novae spectra indeed routinely extend to sufficiently high energies, a much larger fraction of the optically-discovered events are potentially detectable as TeV sources.  Indeed, the total Galactic novae rate is estimated to be $\sim 40$ per year, but the majority go undiscovered due to extinction in the Galactic plane.  Since novae remains gamma-ray luminous for a few weeks, on average at least one nova is gamma-ray active in the Galaxy at any given time.  Novae thus represent a potentially promising transient source for a future CTA survey of the Galactic plane, even in the absence of an optical trigger.

\subsection{Implications for Neutrino Detection}
\label{sec:neutrinos}

An additional prediction of the favored hadronic model is a GeV neutrino flux comparable to the gamma-ray flux that may be detected by future experiments (\citealt{Razzaque+10}). This neutrino energy range is especially interesting with the recent and near-future upgrades of large scale neutrino facilities. Here we consider IceCube with its DeepCore extension \citep{2012APh....35..615A}, which was designed for detection in the $\sim10-100$\,GeV energy range. Future upgrades, such as the Precision IceCube Next Generation Upgrade (PINGU; \citealt{2014arXiv1401.2046T}) or KM3NeT-ORCA \citep{2014arXiv1402.1022K}, can further enhance sensitivity in the relevant energy range.

We begin by considering the detection prospects of novae with IceCube-DeepCore using the observed {\it Fermi} LAT spectrum, which we estimate as $E^{2}dN_{\gamma}/dE \approx 10^{-10} (E/1\,\mbox{GeV})^{-0.1}$erg\,s$^{-1}$cm$^{-2}$ for V1324 Sco 2014 (fairly typical of the LAT-detected novae; \citealt{Ackermann+14}).  In hadronic scenarios, the neutrino energy distribution is approximately equal to the gamma-ray spectrum.  The number of detected neutrinos from nova Sco is therefore approximately given by 
\begin{equation}
N_{\rm \nu} = \int_{10\,\rm GeV}^{E_{\rm max}} \left(\frac{dN_{\gamma}}{dE}\right) A_{\rm eff} T_{\rm sco} dE \approx 0.2 \frac{E_{\rm \nu, max}}{\mbox{TeV}},
\label{eq:Nnu}
\end{equation}
where $T_{\rm sco} \approx 17$ days is the approximate duration of the gamma-ray emission, $E_{\rm \nu, max}$ is the assumed maximum energy of the neutrino energy distribution, and $A_{\rm eff} = 40(E/100\,\mbox{GeV})^2\,$cm$^{2}$ is the IceCube-DeepCore effective area \citep{2012APh....35..615A}.\footnote{This effective area corresponds to the triggering rate of the detector. After applying quality vetoes on these triggers to reduce the background, the effective area is expected to be a factor of a few lower.}

Equation (\ref{eq:Nnu}) shows that the number of detected neutrinos is $\ll 1$ for the spectral cut-offs of $E_{\rm \nu,max} \approx E_{\rm max}/10 \sim 1-10$ GeV suggested by the {\it MAGIC} observations of V339 Del.  However, the number of detected neutrinos is proportional to $E_{\rm max}$ and increases to $\gtrsim 1$ for $E_{\rm \nu, max} \gtrsim 5$ TeV ($E_{\rm max} \gtrsim 50$ TeV).  Detectability also depend on the atmospheric background neutrino flux, which we estimate to be also $\sim 1$ neutrino over the two week duration of typical nova LAT emission, relatively independent of $E_{\rm\nu, max}$.\footnote{The half-sky background rate of IceCube-DeepCore in the energy range $\sim10-100\,$GeV is $\sim10^3$yr$^{-1}$ \citep{2015PhRvD..91g2004A}.  Adopting an angular uncertainty of $\sim 10^{\circ}$ at 50\,GeV as being average for this energy range \citep{2015arXiv150905029A} and considering the two week duration of a nova of known direction, we arrive at a low-energy ($\sim10-100\,$GeV) neutrino background of $N_{\rm \nu,bg,GeV}\sim 1$.  Considering a case with a higher energy cutoff of $E_{\rm max} \gg100\,$GeV, IceCube detects $\sim10^5$ (up-going) neutrino candidates a year over this energy range \citep{2013arXiv1309.6979I}.  For those events detected as tracks, the angular resolution is $\approx1^{\circ}$. The expected high energy background over two weeks is therefore $N_{\rm \nu,bg,TeV}\sim 1$, comparable to that at lower neutrino energies.}

Given the significant background rate and the very high cutoff energy required for $N_{\nu} \gg 1$, detecting a neutrino signal from the current sample of LAT-detected novae seems unlikely.  Detection prospects may, however, be improved by two factors.  First, note that the LAT emission is typically delayed with respect to the peak of the optical emission by a few days \citep{Ackermann+14}, a fact which is puzzling if a significant fraction of the nova optical emission is indeed shock-powered during the LAT phase (\citealt{Metzger+15}). \citet{Metzger+15} hypothesize that this delay is due to the attenuation of the LAT emission by (inelastic) electron scattering, at early times when the shocks occur at the smallest radii and are passing through the densest material.  If this is true, then the true peak of the shock power could be significantly larger, by up to an order of magnitude, than that inferred from the LAT light curve.  To suppress 100 MeV gamma-rays by inelastic scattering requires a Thomson optical $\tau_{T} \sim n \sigma_{T} R_{\rm ej} \gtrsim 100$, i.e. $n \gtrsim 10^{13}$ cm$^{-3}$ for $R_{\rm ej} \sim v_{\rm ej}\times$ 1 day and $v_{\rm ej} \sim$ 1000 km s$^{-1}$.  From equation (\ref{eq:Emax2}), the maximum particle energy is $E_{\rm max} \sim 400{\rm \,GeV}$ for $n = 10^{13}$ cm$^{-3}, v_8 = 1, T_{\rm us} = 2\times 10^{4}$ K and $J_{\nu} \sim 10 J_{\rm f}$.  So, in principle $\sim 1$ TeV energy particles can be accelerated at times when $\gtrsim$ 100 MeV gamma-rays are blocked.


Considering a scenario in which the neutrino luminosity is $\sim 10$ times higher than in our fiducial estimate, but lasting only the first $\sim$ day of the nova outburst, then the total number of detected neutrinos would be similar to our estimate in equation (\ref{eq:Nnu}) (for the same $E_{\rm \nu,max}$), but the neutrino background would be reduced to $N_{\rm \nu, bg} \sim 0.1$ over this shorter time interval.  The prospects for detecting a nova during the earliest phases of the nova outburst would therefore be considerably greater.    

The probability of detecting individual novae would also be higher for a particularly nearby event, e.g. at a distance of 1 kpc instead of the $\sim 4-8$ kpc typical of the current LAT sample.  Finally, given the relatively high rate of Galactic novae ($\sim 10$ yr$^{-1}$), it may be possible to stack multiple novae in one observation, improving the probability of a statistically significant detection.   


\subsection{Magnetic Field Amplification at Non-Relativistic Shocks}
\label{sec:Bfield}

The minimum gamma cut-off of $\sim 3$ GeV allowed by LAT measurements requires particles to be accelerated to $E_{\rm max} \gtrsim 30$ GeV in hadronic scenarios.  From equation (\ref{eq:Emax2}) this requires a minimum magnetic energy fraction of
\begin{eqnarray}
\epsilon_{B} &\gtrsim& 2\times 10^{-3} \epsilon_{\rm p,-1}^{-2}n_{10}v_{8}^{-6}(J_{\nu}/J_{\rm f})^{-2} \nonumber \\
&\approx& 10^{-3}\left(\frac{L_{\gamma}}{10^{36}\rm erg\,s^{-1}}\right)\epsilon_{\rm p,-1}^{-3}\epsilon_{\gamma,-1}^{-1}t_{\rm wk}^{-2}v_{8}^{-11}(J_{\nu}/J_{\rm f})^{-2},
\label{eq:epB}
\end{eqnarray}
where in the final line we have normalized the density using the observed gamma-ray luminosity (eq.~[\ref{eq:Lgamma}]) assuming $R_{\rm ej} \sim v_{\rm sh}t$ and we have assumed that the gamma-ray emission is hadronic.

For $v_{8} \lesssim 1$ this minimum value of $\epsilon_{B}$ is generally larger than the value that would be achieved from flux freezing alone at the shock given the estimated value for the pre-shock field (\citealt{Metzger+15}).  This provides evidence for magnetic field amplification in nova shocks, likely driven by cosmic rays \citep{Bell04}.  There is other evidence for magnetic field amplification in supernova remnant shocks (\citealt{Morlino+10}; \citealt{Ressler+14}), for instance from fitting the combined radio and X-ray data (e.g., \citealt{Morlino&Caprioli12} for Tycho).  However, the evidence for similar amplification is less clear in younger radio supernovae \citep{Thompson+09}.  Novae provide ideal nearby laboratories to study the onset of CR acceleration, because other time-dependent sources, such as radio supernovae, typically occur too distant to detect as gamma-ray sources.

\section{Conclusions}
\label{sec:conclusions}

Shocks in novae occupy a relatively unique regime of low velocities and high gas densities (Fig.~\ref{fig:regimes}).  Such high densities imply that gas well upstream of the shock is neutral and opaque to ionizing radiation.  Because the magnetic field amplification needed for diffusive shock acceleration requires ionized gas, the acceleration process is confined to a narrow photo-ionized layer just ahead of the shock.  This limits particle energies to orders of magnitude below the naive Hillas criterion estimate.  By comparing the downstream advection rate to the growth rate of the NRH Cosmic Ray Current-Driven Instability in the ionized layer (accounting for ion-neutral damping), we have quantified the maximum particle energy, $E_{\rm max}$, across the range of shock velocities and upstream densities constrained independently by observations of nova spectra, characteristic ejecta mass, and LAT gamma-ray emission (Fig.~\ref{fig:Emax1}).  

We find values of $E_{\rm max} \sim 10$ GeV$-$10 TeV, which are broadly consistent with those needed to produce the spectral cut-offs constrained by LAT observations and observed conclusively at a single epoch by the MAGIC telescope.  We also find, however, that in principle acceleration up to $E_{\rm max} \gtrsim 1$ TeV is achievable for low densities and high shock velocities.  Furthermore, the observed spectral cut-offs might not be intrinsic to the accelerator, but instead result from $\gamma-\gamma$ absorption.  The latter is likely to become less important at the highest photon energies (at least until dust formation occurs, producing infrared target photons).  

Novae gamma-ray emission could therefore extend up to the energy range $\gtrsim 100$ GeV accessible to atmosphere Cherenkov telescopes, such as the planned CTA.  Detecting high energy neutrinos by IceCube appears less promising, unless (1) $E_{\rm max} \gtrsim$ 10 TeV and (2) the shock power is higher during the earliest phases of the nova outburst than implied by the LAT emission, or (3) in the case of a particularly nearby nova of distance $\lesssim 1$ kpc, which, however, occur perhaps only once per decade.

Novae are unlikely to be an important source of galactic cosmic rays (CRs) compared to supernova remnants (SNRs).  For canonical Galactic rates of supernovae, $\mathcal R_{\rm SN}\approx 0.03$\,yr$^{-1}$, and novae, $\mathcal R_{\rm N}\approx 40$\,yr$^{-1}$, and blast wave energies of $\mathcal E_{\rm SN}\approx 10^{51}$\,erg and $\mathcal E_{N} \sim M_{\rm ej}v_{\rm ej}^{2}/2 \approx 10^{45}M_{-4}v_{\rm ej,8}^{2}$\,erg, respectively, the relative CR flux ratio is approximately $(\mathcal E_{\rm N}R_{\rm N})/(\mathcal E_{\rm SN}R_{\rm SN})\approx 2\times 10^{-3}$, i.e., the contribution of novae to the diffuse CR spectrum is negligible.  This estimate also assumes equal escape probability, while in fact the nova ejecta is probably more effective than SNRs at trapping cosmic rays due to the higher gas densities and short evolution times.  

Interestingly, the CR spectrum measured by PAMELA and AMS--02 shows a hardening around 200 GeV/nucleon, where the proton spectral slope changes from $\sim$ 2.85 to 2.7 \citep{PAMELA11,AMS15}.  Such a feature could also be explained by an excess of low-energy CRs, whose spectrum is softer than that produced by SNRs (which dominates at larger energies) and cuts off above $\sim$100\,GeV/nucleon.  At first glance, novae represent an intriguing source for this low energy CR component, since their maximum CR energies that fall naturally in the $\sim 100$\,GeV range.  Producing such a large distortion of the CR spectrum would, however, require a comparable nova and SNR contribution to the CR energy flux in this energy range, which as discussed above is only possible in the chance coincidence of a recent very nearby nova.    

The physics of relativistic particle acceleration and magnetic field amplification as probed by nova shocks represent important open problems across many areas of high energy astrophysics.  Our results show that novae provide support for the robustness of both processes in non-relativistic shocks, even under novel conditions of high densities and in the presence of upstream neutral gas.  Given the sensitive dependence of $E_{\rm max}$ on the extent of the ionized layer, future work is needed to better constrain the structure of the upstream gas with a more accurate photo-ionization model, accounting for line emission from behind the radiative shock and also for downscattering and reprocessing of high frequency radiation trapped by the neutral gas.

\section*{Acknowledgments}
BDM gratefully acknowledges support from NASA {\it Fermi} grant NNX14AQ68G, NSF grant AST-1410950, and the Alfred P. Sloan Foundation.  


\begin{thebibliography}{}

\bibitem[\protect\citeauthoryear{{Aartsen}, {Abraham}, {Ackermann}, {Adams},
  {Aguilar}, {Ahlers}, {Ahrens}, {Altmann}, {Anderson}, {Ansseau} \& et
  al.}{{Aartsen} et~al.}{2015}]{2015arXiv150905029A}
{Aartsen} M.~G.,  {Abraham} K.,  {Ackermann} M.,  {Adams} J.,  {Aguilar} J.~A.,
   {Ahlers} M.,  {Ahrens} M.,  {Altmann} D.,  {Anderson} T.,  {Ansseau} I.,
  et al. 2015, ArXiv e-prints

\bibitem[\protect\citeauthoryear{{Aartsen}, {Ackermann}, {Adams}, {Aguilar},
  {Ahlers}, {Ahrens}, {Altmann}, {Anderson}, {Arguelles}, {Arlen} \& et
  al.}{{Aartsen} et~al.}{2015}]{2015PhRvD..91g2004A}
{Aartsen} M.~G.,  {Ackermann} M.,  {Adams} J.,  {Aguilar} J.~A.,  {Ahlers} M.,
  {Ahrens} M.,  {Altmann} D.,  {Anderson} T.,  {Arguelles} C.,  {Arlen} T.~C.,
    et al. 2015, \prd, 91, 072004

\bibitem[\protect\citeauthoryear{{Abbasi}, {Abdou}, {Abu-Zayyad}, {Ackermann},
  {Adams}, {Aguilar}, {Ahlers}, {Allen}, {Altmann}, {Andeen} \& et
  al.}{{Abbasi} et~al.}{2012}]{2012APh....35..615A}
{Abbasi} R.,  {Abdou} Y.,  {Abu-Zayyad} T.,  {Ackermann} M.,  {Adams} J.,
  {Aguilar} J.~A.,  {Ahlers} M.,  {Allen} M.~M.,  {Altmann} D.,  {Andeen} K.,
   et al. 2012, Astroparticle Physics, 35, 615

\bibitem[\protect\citeauthoryear{{Abdo}, {Ackermann}, {Ajello}, {Atwood},
  {Baldini}, {Ballet}, {Barbiellini}, {Bastieri}, {Bechtol}, {Bellazzini} \& et
  al.}{{Abdo} et~al.}{2010}]{Abdo+10}
{Abdo} A.~A.,  {Ackermann} M.,  {Ajello} M.,  {Atwood} W.~B.,  {Baldini} L.,
  {Ballet} J.,  {Barbiellini} G.,  {Bastieri} D.,  {Bechtol} K.,  {Bellazzini}
  R.,    et al. 2010, Science, 329, 817

\bibitem[\protect\citeauthoryear{{Ackermann} et~al.,}{{Ackermann}
  et~al.}{2014}]{Ackermann+14}
{Ackermann} M.,  et~al., 2014, Science, 345, 554

\bibitem[\protect\citeauthoryear{{Actis}, {Agnetta}, {Aharonian},
  {Akhperjanian}, {Aleksi{\'c}}, {Aliu}, {Allan}, {Allekotte}, {Antico},
  {Antonelli} \& et al.}{{Actis} et~al.}{2011}]{Actis+11}
{Actis} M.,  {Agnetta} G.,  {Aharonian} F.,  {Akhperjanian} A.,  {Aleksi{\'c}}
  J.,  {Aliu} E.,  {Allan} D.,  {Allekotte} I.,  {Antico} F.,  {Antonelli}
  L.~A.,    et al. 2011, Experimental Astronomy, 32, 193

\bibitem[\protect\citeauthoryear{{Adriani} et~al.,}{{Adriani}
  et~al.}{2011}]{PAMELA11}
{Adriani} O.,  et~al., 2011, Science, 332, 69

\bibitem[\protect\citeauthoryear{{Aguilar}, {Aisa}, {Alpat}, {Alvino},
  {Ambrosi}, {Andeen}, {Arruda}, {Attig}, {Azzarello}, {Bachlechner} \& et
  al.}{{Aguilar} et~al.}{2015}]{AMS15}
{Aguilar} M.,  {Aisa} D.,  {Alpat} B.,  {Alvino} A.,  {Ambrosi} G.,  {Andeen}
  K.,  {Arruda} L.,  {Attig} N.,  {Azzarello} P.,  {Bachlechner} A.,    et al.
  2015, Physical Review Letters, 114, 171103

\bibitem[\protect\citeauthoryear{{Aliu} et~al.,}{{Aliu}
  et~al.}{2012}]{Aliu+12}
{Aliu} E.,  et~al., 2012, \apj, 754, 77

\bibitem[\protect\citeauthoryear{{Bath} \& {Shaviv}}{{Bath} \&
  {Shaviv}}{1976}]{Bath&Shaviv76}
{Bath} G.~T.,  {Shaviv} G.,  1976, \mnras, 175, 305

\bibitem[\protect\citeauthoryear{{Bell}}{{Bell}}{2004}]{Bell04}
{Bell} A.~R.,  2004, \mnras, 353, 550

\bibitem[\protect\citeauthoryear{{Bell}}{{Bell}}{2005}]{Bell05}
{Bell} A.~R.,  2005, \mnras, 358, 181

\bibitem[\protect\citeauthoryear{{Beloborodov}}{{Beloborodov}}{2014}]{Beloborodov14}
{Beloborodov} A.~M.,  2014, \mnras, 438, 169

\bibitem[\protect\citeauthoryear{{Blandford} \& {Ostriker}}{{Blandford} \&
  {Ostriker}}{1978}]{Blandford&Ostriker78}
{Blandford} R.~D.,  {Ostriker} J.~P.,  1978, \apjl, 221, L29

\bibitem[\protect\citeauthoryear{{Blasi} \& {Amato}}{{Blasi} \&
  {Amato}}{2008}]{Amato&Blasi08}
{Blasi} P.,  {Amato} E.,  2008, International Cosmic Ray Conference, 2, 235

\bibitem[\protect\citeauthoryear{{Blasi} \& {Amato}}{{Blasi} \&
  {Amato}}{2012}]{ba12}
{Blasi} P.,  {Amato} E.,  2012, \jcap, 1, 11

\bibitem[\protect\citeauthoryear{{Blasi}, {Morlino}, {Bandiera}, {Amato} \&
  {Caprioli}}{{Blasi} et~al.}{2012}]{Blasi+12}
{Blasi} P.,  {Morlino} G.,  {Bandiera} R.,  {Amato} E.,    {Caprioli} D.,
  2012, \apj, 755, 121

\bibitem[\protect\citeauthoryear{{Caprioli} \& {Spitkovsky}}{{Caprioli} \&
  {Spitkovsky}}{2014a}]{Caprioli&Spitkovsky14}
{Caprioli} D.,  {Spitkovsky} A.,  2014a, \apj, 783, 91

\bibitem[\protect\citeauthoryear{{Caprioli} \& {Spitkovsky}}{{Caprioli} \&
  {Spitkovsky}}{2014b}]{Caprioli&Spitkovsky14b}
{Caprioli} D.,  {Spitkovsky} A.,  2014b, \apj, 794, 46

\bibitem[\protect\citeauthoryear{{Caprioli} \& {Spitkovsky}}{{Caprioli} \&
  {Spitkovsky}}{2014c}]{Caprioli&Spitkovsky14c}
{Caprioli} D.,  {Spitkovsky} A.,  2014c, \apj, 794, 47

\bibitem[\protect\citeauthoryear{{Chevalier} \& {Irwin}}{{Chevalier} \&
  {Irwin}}{2011}]{Chevalier&Irwin11}
{Chevalier} R.~A.,  {Irwin} C.~M.,  2011, \apjl, 729, L6

\bibitem[\protect\citeauthoryear{{Chomiuk}, {Linford}, {Yang}, {O'Brien},
  {Paragi}, {Mioduszewski}, {Beswick}, {Cheung}, {Mukai}, {Nelson}, {Ribeiro},
  {Rupen}, {Sokoloski}, {Weston}, {Zheng}, {Bode}, {Eyres}, {Roy} \&
  {Taylor}}{{Chomiuk} et~al.}{2014}]{Chomiuk+14}
{Chomiuk} L.,  {Linford} J.~D.,  {Yang} J.,  {O'Brien} T.~J.,  {Paragi} Z.,
  {Mioduszewski} A.~J.,  {Beswick} R.~J.,  {Cheung} C.~C.,  {Mukai} K.,
  {Nelson} T.,  {Ribeiro} V.~A.~R.~M.,  {Rupen} M.~P.,  {Sokoloski} J.~L.,
  {Weston} J.,  {Zheng} Y.,  {Bode} M.~F.,  {Eyres} S.,  {Roy} N.,    {Taylor}
  G.~B.,  2014, \nat, 514, 339

\bibitem[\protect\citeauthoryear{{Finzell}, {Chomiuk}, {Munari} \&
  {Walter}}{{Finzell} et~al.}{2015}]{Finzell+15}
{Finzell} T.,  {Chomiuk} L.,  {Munari} U.,    {Walter} F.~M.,  2015, \apj, 809,
  160

\bibitem[\protect\citeauthoryear{{Hillas}}{{Hillas}}{1984}]{Hillas+84}
{Hillas} A.~M.,  1984, \araa, 22, 425

\bibitem[\protect\citeauthoryear{{Hillman}, {Prialnik}, {Kovetz}, {Shara} \&
  {Neill}}{{Hillman} et~al.}{2014}]{Hillman+14}
{Hillman} Y.,  {Prialnik} D.,  {Kovetz} A.,  {Shara} M.~M.,    {Neill} J.~D.,
  2014, \mnras, 437, 1962

\bibitem[\protect\citeauthoryear{{IceCube Collaboration}, {Aartsen}, {Abbasi},
  {Abdou}, {Ackermann}, {Adams}, {Aguilar}, {Ahlers}, {Altmann}, {Auffenberg}
  \& et al.}{{IceCube Collaboration} et~al.}{2013}]{2013arXiv1309.6979I}
{IceCube Collaboration} {Aartsen} M.~G.,  {Abbasi} R.,  {Abdou} Y.,
  {Ackermann} M.,  {Adams} J.,  {Aguilar} J.~A.,  {Ahlers} M.,  {Altmann} D.,
  {Auffenberg} J.,    et al. 2013, ArXiv e-prints

\bibitem[\protect\citeauthoryear{{Kamae}, {Karlsson}, {Mizuno}, {Abe} \&
  {Koi}}{{Kamae} et~al.}{2006}]{Kamae+06}
{Kamae} T.,  {Karlsson} N.,  {Mizuno} T.,  {Abe} T.,    {Koi} T.,  2006, \apj,
  647, 692

\bibitem[\protect\citeauthoryear{{Karle} et~al.,}{{Karle}
  et~al.}{2003}]{Karle+03}
{Karle} A.,  et~al., 2003, Nuclear Physics B Proceedings Supplements, 118, 388

\bibitem[\protect\citeauthoryear{{Kato}}{{Kato}}{2014}]{Kato14}
{Kato} T.~N.,  2014, ArXiv e-prints

\bibitem[\protect\citeauthoryear{{Katz}}{{Katz}}{2014}]{2014arXiv1402.1022K}
{Katz} U.~F.,  2014, ArXiv e-prints

\bibitem[\protect\citeauthoryear{{Kulsrud} \& {Cesarsky}}{{Kulsrud} \&
  {Cesarsky}}{1971}]{Kulsrud&Cesarsky71}
{Kulsrud} R.~M.,  {Cesarsky} C.~J.,  1971, ApJ Letters, 8, 189

\bibitem[\protect\citeauthoryear{{Livio}, {Shankar}, {Burkert} \&
  {Truran}}{{Livio} et~al.}{1990}]{Livio+90}
{Livio} M.,  {Shankar} A.,  {Burkert} A.,    {Truran} J.~W.,  1990, \apj, 356,
  250

\bibitem[\protect\citeauthoryear{{Lloyd}, {O'Brien} \& {Bode}}{{Lloyd}
  et~al.}{1997}]{Lloyd+97}
{Lloyd} H.~M.,  {O'Brien} T.~J.,    {Bode} M.~F.,  1997, \mnras, 284, 137

\bibitem[\protect\citeauthoryear{{MAGIC Collaboration}, {Ahnen}, {Ansoldi},
  {Antonelli} et~al.,}{{MAGIC Collaboration} et~al.}{2015}]{MAGIC+15}
{MAGIC Collaboration} {Ahnen} M.~L.,  {Ansoldi} S.,  {Antonelli} L.~A.,
  et~al., 2015, ArXiv e-prints

\bibitem[\protect\citeauthoryear{{Martin} \& {Dubus}}{{Martin} \&
  {Dubus}}{2013}]{Martin&Dubus13}
{Martin} P.,  {Dubus} G.,  2013, \aap, 551, A37

\bibitem[\protect\citeauthoryear{{Metzger}, {Finzell}, {Vurm}, {Hasco{\"e}t},
  {Beloborodov} \& {Chomiuk}}{{Metzger} et~al.}{2015}]{Metzger+15}
{Metzger} B.~D.,  {Finzell} T.,  {Vurm} I.,  {Hasco{\"e}t} R.,  {Beloborodov}
  A.~M.,    {Chomiuk} L.,  2015, \mnras, 450, 2739

\bibitem[\protect\citeauthoryear{{Metzger}, {Hasco{\"e}t}, {Vurm},
  {Beloborodov}, {Chomiuk}, {Sokoloski} \& {Nelson}}{{Metzger}
  et~al.}{2014}]{Metzger+14}
{Metzger} B.~D.,  {Hasco{\"e}t} R.,  {Vurm} I.,  {Beloborodov} A.~M.,
  {Chomiuk} L.,  {Sokoloski} J.~L.,    {Nelson} T.,  2014, \mnras, 442, 713

\bibitem[\protect\citeauthoryear{{Morlino}, {Amato}, {Blasi} \&
  {Caprioli}}{{Morlino} et~al.}{2010}]{Morlino+10}
{Morlino} G.,  {Amato} E.,  {Blasi} P.,    {Caprioli} D.,  2010, \mnras, 405,
  L21

\bibitem[\protect\citeauthoryear{{Morlino} \& {Caprioli}}{{Morlino} \&
  {Caprioli}}{2012}]{Morlino&Caprioli12}
{Morlino} G.,  {Caprioli} D.,  2012, \aap, 538, A81

\bibitem[\protect\citeauthoryear{{Mukai} \& {Ishida}}{{Mukai} \&
  {Ishida}}{2001}]{Mukai&Ishida01}
{Mukai} K.,  {Ishida} M.,  2001, \apj, 551, 1024

\bibitem[\protect\citeauthoryear{{Mukai}, {Orio} \& {Della Valle}}{{Mukai}
  et~al.}{2008}]{Mukai+08}
{Mukai} K.,  {Orio} M.,    {Della Valle} M.,  2008, \apj, 677, 1248

\bibitem[\protect\citeauthoryear{{Osborne}}{{Osborne}}{2015}]{Osborne+15}
{Osborne} J.~P.,  2015, Journal of High Energy Astrophysics, 7, 117

\bibitem[\protect\citeauthoryear{{Osterbrock} \& {Ferland}}{{Osterbrock} \&
  {Ferland}}{2006}]{Osterbrock&Ferland06}
{Osterbrock} D.~E.,  {Ferland} G.~J.,  2006, {Astrophysics of gaseous nebulae
  and active galactic nuclei}

\bibitem[\protect\citeauthoryear{{Park}, {Caprioli} \& {Spitkovsky}}{{Park}
  et~al.}{2014}]{Park+14}
{Park} J.,  {Caprioli} D.,    {Spitkovsky} A.,  2014, ArXiv e-prints

\bibitem[\protect\citeauthoryear{{Pinto} \& {Eastman}}{{Pinto} \&
  {Eastman}}{2000}]{Pinto&Eastman00}
{Pinto} P.~A.,  {Eastman} R.~G.,  2000, \apj, 530, 757

\bibitem[\protect\citeauthoryear{{Razzaque}, {Jean} \& {Mena}}{{Razzaque}
  et~al.}{2010}]{Razzaque+10}
{Razzaque} S.,  {Jean} P.,    {Mena} O.,  2010, \prd, 82, 123012

\bibitem[\protect\citeauthoryear{{Ressler}, {Katsuda}, {Reynolds}, {Long},
  {Petre}, {Williams} \& {Winkler}}{{Ressler} et~al.}{2014}]{Ressler+14}
{Ressler} S.~M.,  {Katsuda} S.,  {Reynolds} S.~P.,  {Long} K.~S.,  {Petre} R.,
  {Williams} B.~J.,    {Winkler} P.~F.,  2014, \apj, 790, 85

\bibitem[\protect\citeauthoryear{{Reville}, {Kirk} \& {Duffy}}{{Reville}
  et~al.}{2006}]{Reville+06}
{Reville} B.,  {Kirk} J.~G.,    {Duffy} P.,  2006, Plasma Physics and
  Controlled Fusion, 48, 1741

\bibitem[\protect\citeauthoryear{{Reville}, {Kirk}, {Duffy} \&
  {O'Sullivan}}{{Reville} et~al.}{2007}]{Reville+07}
{Reville} B.,  {Kirk} J.~G.,  {Duffy} P.,    {O'Sullivan} S.,  2007, \aap, 475,
  435

\bibitem[\protect\citeauthoryear{{Ribeiro}, {Munari} \& {Valisa}}{{Ribeiro}
  et~al.}{2013}]{Ribeiro+13}
{Ribeiro} V.~A.~R.~M.,  {Munari} U.,    {Valisa} P.,  2013, \apj, 768, 49

\bibitem[\protect\citeauthoryear{{Riquelme} \& {Spitkovsky}}{{Riquelme} \&
  {Spitkovsky}}{2009}]{Riquelme&Spitkovsky09}
{Riquelme} M.~A.,  {Spitkovsky} A.,  2009, \apj, 694, 626

\bibitem[\protect\citeauthoryear{{Schaefer} et~al.,}{{Schaefer}
  et~al.}{2014}]{Schaefer+14}
{Schaefer} G.~H.,  et~al., 2014, \nat, 515, 234

\bibitem[\protect\citeauthoryear{{Schure}, {Kosenko}, {Kaastra}, {Keppens} \&
  {Vink}}{{Schure} et~al.}{2009}]{Schure+09}
{Schure} K.~M.,  {Kosenko} D.,  {Kaastra} J.~S.,  {Keppens} R.,    {Vink} J.,
  2009, \aap, 508, 751

\bibitem[\protect\citeauthoryear{{Schwarz}, {Starrfield}, {Shore} \&
  {Hauschildt}}{{Schwarz} et~al.}{1997}]{Schwarz+97}
{Schwarz} G.~J.,  {Starrfield} S.,  {Shore} S.~N.,    {Hauschildt} P.~H.,
  1997, \mnras, 290, 75

\bibitem[\protect\citeauthoryear{{Seaquist}, {Duric}, {Israel}, {Spoelstra},
  {Ulich} \& {Gregory}}{{Seaquist} et~al.}{1980}]{Seaquist+80}
{Seaquist} E.~R.,  {Duric} N.,  {Israel} F.~P.,  {Spoelstra} T.~A.~T.,  {Ulich}
  B.~L.,    {Gregory} P.~C.,  1980, \aj, 85, 283

\bibitem[\protect\citeauthoryear{{Shore}}{{Shore}}{2012}]{Shore12}
{Shore} S.~N.,  2012, Bulletin of the Astronomical Society of India, 40, 185

\bibitem[\protect\citeauthoryear{{Shore}, {De Gennaro Aquino}, {Schwarz},
  {Augusteijn}, {Cheung}, {Walter} \& {Starrfield}}{{Shore}
  et~al.}{2013}]{Shore+13}
{Shore} S.~N.,  {De Gennaro Aquino} I.,  {Schwarz} G.~J.,  {Augusteijn} T.,
  {Cheung} C.~C.,  {Walter} F.~M.,    {Starrfield} S.,  2013, \aap, 553, A123

\bibitem[\protect\citeauthoryear{{Starrfield}, {Truran}, {Wiescher} \&
  {Sparks}}{{Starrfield} et~al.}{1998}]{Starrfield+98}
{Starrfield} S.,  {Truran} J.~W.,  {Wiescher} M.~C.,    {Sparks} W.~M.,  1998,
  \mnras, 296, 502

\bibitem[\protect\citeauthoryear{{The IceCube-PINGU Collaboration}}{{The
  IceCube-PINGU Collaboration}}{2014}]{2014arXiv1401.2046T}
{The IceCube-PINGU Collaboration} 2014, ArXiv e-prints

\bibitem[\protect\citeauthoryear{{Thompson}, {Quataert} \& {Murray}}{{Thompson}
  et~al.}{2009}]{Thompson+09}
{Thompson} T.~A.,  {Quataert} E.,    {Murray} N.,  2009, \mnras, 397, 1410

\bibitem[\protect\citeauthoryear{{V{\"o}lk}, {Berezhko} \&
  {Ksenofontov}}{{V{\"o}lk} et~al.}{2005}]{Volk+05}
{V{\"o}lk} H.~J.,  {Berezhko} E.~G.,    {Ksenofontov} L.~T.,  2005, \aap, 433,
  229

\bibitem[\protect\citeauthoryear{{Weston}, {Sokoloski}, {Metzger}, {Zheng},
  {Chomiuk}, {Krauss}, {Linford}, {Nelson}, {Mioduszewski}, {Rupen}, {Finzell}
  \& {Mukai}}{{Weston} et~al.}{2015}]{Weston+15}
{Weston} J.~H.~S.,  {Sokoloski} J.~L.,  {Metzger} B.~D.,  {Zheng} Y.,
  {Chomiuk} L.,  {Krauss} M.~I.,  {Linford} J.,  {Nelson} T.,  {Mioduszewski}
  A.,  {Rupen} M.~P.,  {Finzell} T.,    {Mukai} K.,  2015, ArXiv e-prints

\bibitem[\protect\citeauthoryear{{Williams} \& {Mason}}{{Williams} \&
  {Mason}}{2010}]{Williams&Mason10}
{Williams} R.,  {Mason} E.,  2010, \apss, 327, 207

\bibitem[\protect\citeauthoryear{{Zirakashvili}, {Ptuskin} \&
  {V{\"o}lk}}{{Zirakashvili} et~al.}{2008}]{Zirakashvili+08}
{Zirakashvili} V.~N.,  {Ptuskin} V.~S.,    {V{\"o}lk} H.~J.,  2008, \apj, 678,
  255

\end{thebibliography}


\end{document}